\documentclass[12pt,a4paper,twoside]{article}

\usepackage[applemac]{inputenc}
\usepackage[english]{babel}
\usepackage[T1]{fontenc}
\usepackage{lmodern}
\usepackage{geometry} 
\usepackage[intlimits,sumlimits]{amsmath}
\usepackage{amssymb}
\usepackage[amsthm,thmmarks]{ntheorem}
\usepackage{mathrsfs}
\usepackage{bm}
\usepackage{xpatch}
\usepackage{dsfont}
\usepackage{siunitx} 
\usepackage{scalerel,stackengine}
\usepackage{aligned-overset}
\usepackage{algorithm}
\usepackage{algpseudocode}
\usepackage{multicol}
\usepackage{booktabs}
\usepackage{multirow}
\usepackage{makecell}

\usepackage[backend=bibtex, style=authoryear, firstinits=true, maxcitenames=2, sorting=nty, isbn=false, dashed=false]{biblatex}
\DeclareNameAlias{sortname}{last-first}
\DeclareFieldFormat*{title}{#1}
\DeclareFieldFormat*{edition}{#1 edition}
\DeclareDelimFormat[cite,parencite]{nameyeardelim}{\textcolor{black}{\addcomma\space}}

\DeclareFieldFormat{citehyperref}{%
   \DeclareFieldAlias{bibhyperref}{noformat}
   \bibhyperref{#1}}

\savebibmacro{cite}

\renewbibmacro*{cite}{%
   \printtext[citehyperref]{%
      \restorebibmacro{cite}%
      \usebibmacro{cite}}}
\usepackage{xpatch}
\xpatchbibmacro{textcite}
  {\printnames{labelname}}
  {\printtext[bibhyperref]{\printnames{labelname}}}
  {}{}

\usepackage{caption}
\usepackage{subcaption}
\usepackage{paralist}
\usepackage{graphicx}
\usepackage{colortbl}
\usepackage[percent]{overpic}
\usepackage{mathtools}
\usepackage[usenames,dvipsnames,svgnames,table]{xcolor}
\usepackage{pgf,tikz}
\usetikzlibrary{patterns}
\usetikzlibrary{quotes,babel,angles}
\usetikzlibrary{arrows}
\usepackage{rotating}
\usepackage{pgfplots}
\usepackage{wrapfig}
\usepackage[most]{tcolorbox}
\usepackage[compact]{titlesec}
\usepackage{fancyhdr}

\definecolor{UHwhite}{RGB}{255,255,255}
\definecolor{UHorange}{RGB}{255,128,0}
\definecolor{UHred}{RGB}{202,53,56}
\definecolor{UHgruen}{RGB}{161,176,45}
\definecolor{UHdarkblue}{RGB}{6,68,107}
\definecolor{UHblue}{RGB}{24,105,183}
\definecolor{UHmidblue}{RGB}{17,119,182}
\definecolor{UHcyan}{RGB}{87,168,211}
\definecolor{UHbrightblue}{RGB}{160,206,234}
\definecolor{UHdarkgrey}{RGB}{69,69,69}
\definecolor{UHgrey}{RGB}{148,148,148}
\definecolor{UHbrightgrey}{RGB}{215,215,215}
\definecolor{UHbeige}{RGB}{235,230,215}

\definecolor{myred}{RGB}{178,34,34}
\definecolor{mygrey}{RGB}{69,69,69}

\numberwithin{equation}{section}
\definecolor{blau}{RGB}{24,105,183}
\usepackage[colorlinks=False,
pdfpagelabels,
pdfstartview = FitH,
bookmarksopen = true,
bookmarksnumbered = true,
colorlinks   = true,
linkcolor = black,
citecolor = UHblue,
plainpages = false,
hypertexnames = false,
linkbordercolor = UHblue,
urlcolor     = UHblue,
citebordercolor=white,
urlbordercolor=white] {hyperref}
\usepackage{orcidlink}

\usepackage{sectsty}
\usepackage{booktabs}

\usepackage[justification=centering]{caption}
\usepackage{dcolumn}
\usepackage{framed}
\newcolumntype{2}{D{.}{}{3.0}}
\makeatletter 
\renewcommand*{\@pnumwidth}{3em}      
\makeatother
\makeatletter
\renewcommand{\p@envcount}{\thesubsection.}
\makeatother

\newcommand{\R}{\mathbb{R}}

\newcommand{\mng}[1]{\left \{ #1 \right \}}

\newcommand{\spn}[1]{\mathrm{span}\kern-0.4ex\left(#1\right)}
\newcommand{\supp}[1]{\mathrm{supp}\kern-0.4ex\left(#1\right)}
\newcommand{\bild}[1]{\mathrm{Bild}\kern-0.4ex\left( #1\right)}
\newcommand{\krn}[1]{\mathrm{Kern}\kern-0.4ex\left( #1\right)}
\newcommand{\rang}[1]{\mathrm{rang}\kern-0.4ex\left( #1\right)}
\newcommand{\spur}[1]{\mathrm{Spur}\kern-0.4ex\left( #1\right)}

\newcommand{\ggt}[1]{\mathrm{ggT}\kern-0.4ex\left( #1\right)}
\newcommand{\kgv}[1]{\mathrm{kgV}\kern-0.4ex\left( #1\right)}

\newcommand{\grpGL}[1]{\mathrm{GL}\kern-0.4ex\left( #1\right)}
\newcommand{\grpSL}[1]{\mathrm{SL}\kern-0.4ex\left( #1\right)}
\newcommand{\grpO}[1]{\mathrm{O}\kern-0.4ex\left( #1\right)}
\newcommand{\grpSO}[1]{\mathrm{SO}\kern-0.4ex\left( #1\right)}
\newcommand{\grpU}[1]{\mathrm{U}\kern-0.4ex\left( #1\right)}
\newcommand{\grpSU}[1]{\mathrm{SU}\kern-0.4ex\left( #1\right)}

\renewcommand{\P}[1]{\mathbb{P}\kern-0.4ex\left( #1\right)}
\newcommand{\E}[1]{\mathbb{E}\kern-0.4ex\left( #1\right)}
\DeclareMathOperator*{\argmax}{arg\,max}
\DeclareMathOperator*{\argmin}{arg\,min}

\makeatletter
\renewenvironment{abstract}{%
    \if@twocolumn
      \section*{\abstractname}%
    \else 
      \begin{center}%
        {\bfseries \abstractname\vspace{\z@}}%
      \end{center}%
      \quotation
    \fi}
    {\if@twocolumn\else\endquotation\fi}
\makeatother

\begin{document}
\sloppy


\title{\fontsize{24}{30} \selectfont \textbf{Gradient-Boosted Mixture Regression Models for Postprocessing\\ Ensemble Weather Forecasts}}

\author{David Jobst\,\orcidlink{0000-0002-2014-3530}\thanks{Corresponding author, University of Hildesheim, Institute of Mathematics and Applied Informatics, Samelsonplatz 1, 31141 Hildesheim, Germany, \texttt{\href{mailto:jobstd@uni-hildesheim.de}{jobstd@uni-hildesheim.de}}}}
\maketitle
\thispagestyle{empty}

\begin{abstract}
\small Nowadays, weather forecasts are commonly generated by ensemble forecasts based on multiple runs of numerical weather prediction models. However, such forecasts are usually miscalibrated and/or biased, thus require statistical postprocessing. Non-homogeneous regression models, such as the ensemble model output statistics are frequently applied to correct these forecasts. Nonetheless, these methods often rely on the assumption of an unimodal parametric distribution, leading to improved, but sometimes not fully calibrated forecasts. To address this issue, a mixture regression model is presented, where the ensemble forecasts of each exchangeable group are linked to only one mixture component and mixture weight, called mixture of model output statistics (MIXMOS). In order to remove location specific effects and to use a longer training data, the standardized anomalies of the response and the ensemble forecasts are employed for the mixture of standardized anomaly model output statistics (MIXSAMOS). As carefully selected covariates, e.g. from different weather variables, can enhance model performance, the non-cyclic gradient-boosting algorithm for mixture regression models is introduced. Furthermore, MIXSAMOS is extended by this gradient-boosting algorithm (MIXSAMOS-GB) providing an automatic variable selection. The novel mixture regression models substantially outperform state-of-the-art postprocessing models in a case study for 2\,\si{m} surface temperature forecasts in Germany.

\end{abstract}

\newpage

\section{Introduction}
\label{sec: Introduction}

Since the first operational runs of ensemble prediction systems (EPSs) at the end of 1992 by the European Centre for Medium-Range Weather Forecasts (ECMWF, \cites{Buizza1993, Molteni1996}) and the National Centers for Environmental Predicition (NCEP, \cites{Toth1993, Tracton1993}), probabilistic weather forecasting employing ensembles of forecasts has become state-of-the-art. An EPS is based on a numerical weather prediction (NWP) model which is run multiple times with different perturbed initial conditions and/or model configurations in order to generate an ensemble of forecasts \parencite{Leutbecher2008}. The EPS based on the Integrated Forecasting System (IFS) of the ECMWF generates, for example, a single control forecast using the best guess of the initial state of the atmosphere. Additionally, it produces 50 perturbed ensemble forecasts obtained by slightly perturbed initial states from the one used for the control forecast \parencite{ECMWF2012}. Although the EPSs are continuously improved, the ensemble forecasts still suffer from systematic bias and/or underdispersion. Therefore, in order to obtain unbiased and calibrated weather predictions, these ensemble forecasts require statistical postprocessing in coherence with recent observations \parencite{Vannitsem2021}.  

One of the first and still very popular parametric postprocessing techniques is the ensemble model output statistics (EMOS, \cite{Gneiting2005}). This distributional regression approach \parencite{Kneib2023} employs the ensemble forecasts as covariates within the linear predictors for each distribution parameter. If different types of forecasts or multi-model ensemble forecasts are available, groups of statistically exchangeable \parencite{Kingman1978} ensemble forecasts, can be identified. Consequently, the regression coefficients for all forecast members within an exchangeable group are constraint to be the same \parencite{Gneiting2014b}. Building upon the EMOS framework, \textcite{Dabernig2017} suggest the standardized anomaly model output statistics (SAMOS) method. In contrast to EMOS, this method utilizes the standardized anomalies of the observations and ensemble forecast by removing station specific characteristics such as seasonal effects. Furthermore, SAMOS allows to integrate groups of exchangeable ensemble forecasts in the same manner as EMOS yielding statistically consistent predictions. Eventually, SAMOS employs by design more training data than the original EMOS approach, which is one potential reason for its improvements over EMOS \parencite{Lang2020}.

However, unimodal parametric distributions are typically assumed to model the weather variable  of interest within the EMOS or SAMOS framework, e.g. the normal or logistic distribution for temperature \parencite{Gneiting2005, Gebetsberger2019} and adaptions thereof such as the truncated normal or log-normal distribution for wind speed \parencites{Thorarinsdottir2010, Baran2015a} and censored logistic distribution for precipitation \parencite{Gebetsberger2016, Stauffer2017}. Although, both frameworks allow to include exchangeable groups of ensemble forecasts, additional information contained in these forecasts remains mostly undetected. Firstly, the forecast ensemble might be multimodal which can not be reflected by an unimodal predictive distribution. Secondly, shape characteristics of the raw ensemble such as kurtosis or skewness remain ignored, if not explicitly modeled, see, e.g. \textcites{Gebetsberger2019, Allen2021}. However, taking account of such properties could improve the calibration and the overall performance of the forecasts.

One possibility to overcome these limitations, is the use of (finite) mixture regression models \parencite{Quandt1972}. This is a distributional regression approach, where the predictive probability density function (PDF) is given by a weighted sum of PDFs (mixture components), where each distribution parameter is allowed to depend on covariates \parencite{Kneib2023}. The Bayesian model averaging (BMA, \cite{Raftery2005}) approach is an example for such a mixture regression model frequently applied in the context of ensemble postprocessing, see, e.g. \textcites{Sloughter2007, Roquelaure2008, Sloughter2010, Bao2010}. For each mixture component, the same parametric distribution family is usually assumed and each forecast member serves as covariate for one or sometimes more distribution parameters of a single mixture component. To account for exchangeability in BMA, \textcite{Fraley2010} suggest to use the same mixture weights and regression coefficients for all members within one group similar to the EMOS setting. Although it is theoretically feasible to estimate all distribution parameters jointly, in practice, BMA is frequently estimated using a step-by-step approach. A different mixture regression approach has been suggested by \textcite{Baran2016}, where a weighted mixture of EMOS models based on different parametric distribution families for each EMOS mixture component is estimated in one step. If exchangeable groups are available, the distribution parameters are parameterized in the same manner akin to EMOS, which is essentially different to BMA. 
Initially tested for wind speed forecasts, this approach has been extended for postprocessing precipitation \parencite{Baran2018} and recently visibility \parencite{Baran2024} forecasts. 
In comparison to BMA and the weighted mixture of EMOS models, \textcite{Taillardat2021} suggests another type of mixture regression model, where the ensemble forecasts of each exchangeable group correspond to only one mixture component, yielding calibrated temperature forecasts. Secondly, different distribution families can be assumed among the mixture components, allowing a flexible distribution modeling. Eventually, the model is estimated in a single step to avoid overdispersive models, which typically arises from two-step approaches such as the linear (opinion) pool \parencite{Stone1961}.

The previously presented mixture regression models mostly include only ensemble forecast of the weather variable under consideration and no further predictor variables. Exceptions are the works of \textcites{Chmielecki2011, Eide2017}, who integrated ensemble forecasts of a small number of hand selected weather variables within the BMA framework to postprocess visibility and wind speed forecasts, respectively. According to the authors, these auxiliary predictor variables enhance the model performance. However, the selection of relevant covariates from a large set including e.g. ensemble forecasts of different weather quantities, lead times, types and locations as well as (lagged) observations or interactions thereof requires expertise. Furthermore, there is still the possibility to select too many variables resulting in multicollineartiy issues and/or overfitting. To handle such a situation within the SAMOS framework, \textcite{Messner2017} has extended the SAMOS model by a gradient-boosting (GB) algorithm, called SAMOS-GB in the following. By construction of the estimation procedure, relevant covariates are included into the model and non-relevant covariates are ignored, which often yields to a superior predictive model performance. 

However, to the best of the authors knowledge, no mixture regression model supporting an automatic covariate selection has been investigated within the field of ensemble postprocessing, yet. Therefore, this research gap identified by \textcite{Eide2017} is  addressed by a mixture regression model, which is estimated via gradient-boosting. This approach allows to estimate a flexible predictive distribution in combination with an explainable variable selection for all distribution parameters. In more detail, the contribution of the paper is the following:

\begin{compactenum}[(i)]
    \item The author proposes mixture regression models for postprocessing ensemble weather forecasts in the fashion of \textcites{Taillardat2021}, where each mixture component corresponds to a single exchangeable group. Furthermore, not only the parameters of each mixture component but also its corresponding weights are allowed to depend on covariates, now. As such a mixture regression model is a \textit{mixture of model output statistics}, it is called MIXMOS. If standardized anomalies of the observations and ensemble forecasts are used for this mixture regression model, it is named \textit{mixture of standardized anomaly model output statistics} (MIXSAMOS).
    \item Based on the initial ideas of gradient-boosted mixture regression models by \textcite{Hepp2023}, a non-cyclic gradient-boosting estimation algorithm \parencite{Thomas2017} for the general class of mixture regression models is presented and also implemented in the \texttt{R}-package \texttt{mixnhreg} \parencite{Jobst2024c}. Besides the typically used logarithmic score, the continuous ranked probability score is considered as additional loss function. In the application context, this gradient-boosting estimation algorithm is employed to estimate MIXSAMOS, giving the name MIXSAMOS-GB. 
    \item MIXSAMOS and its gradient-boosting extension MIXSAMOS-GB are applied in a case study for postprocessing 2\,\si{m} surface temperature forecasts at 280 observation stations in Germany. Both approaches are compared to the state-of-the-art postprocessing methods SAMOS and SAMOS-GB. The results show, that MIXSAMOS performs sightly better than SAMOS with respect to the considered scores, while MIXSAMOS-GB significantly outperforms all other methods. Furthermore, MIXSAMOS-GB provides informative insights in the covariate selection for all distribution parameters. 
\end{compactenum}

The remainder of the paper is organized as follows: Section \ref{sec: Mixture regression models} introduces the general setting for mixture regression models and its non-cyclic gradient-boosting estimation algorithm. The data as well as its usage for the case study including some notation are presented in section \ref{sec: Data}. In section \ref{sec: Ensemble postprocessing methods}, the benchmark methods SAMOS and SAMOS-GB are explained. In addition, the novel mixture regression models designed specifically for the ensemble postprocessing context $-$ namely, MIXMOS, MIXSAMOS, and MIXSAMOS-GB $-$ are introduced. Forecast evaluation methods for model comparison are summarized in section \ref{sec: Forecast evaluation}. The results of the case study are discussed in section \ref{sec: Results}. The paper closes with a conclusion and outlook in section \ref{sec: Conclusion and outlook}.

\section{Mixture regression models}
\label{sec: Mixture regression models}

This section starts with a short introduction to (finite) mixture regression model. Afterwards, the non-cyclic gradient-boosting algorithm is presented for this model class.

\subsection{General model definition}
\label{sec: General model definition}

Finite mixture models are employed to model the distribution of pooled observations arising by unknown proportions from a finite number of unobserved components (classes). Consequently, the distribution of a mixture model is given by a weighted sum of PDFs, where the mixture weights represent the (unknown) proportions and the PDFs stand for the (unobserved) mixture components \parencite{Lindsay1995}. Eventually, a central goal of mixture modeling is to specify and estimate all mixture weights and mixture components suitably. Due to its flexibility, mixture models have been extensively used e.g. for classifying observations, accounting for clustering, and modeling unobserved heterogeneity. For a short overview concerning mixture models, see, e.g. \textcite{McLachlan2019} and for a more detailed introduction have a look at e.g. \textcite{McLachlan2000}. 

Mixture regression models build upon the framework of  mixture models, with the goal to explain the conditional distribution of a response given a set of covariates. This extension has been first proposed by \textcite{Quandt1972} and refined by \textcites{DeSarbo1988, Wedel1995}, who linked covariates in terms of a generalized linear model (GLM, \cite{Nelder1972}) to the mean parameters of more than two mixture components. This methodology enables the classification of a sample into various groups and at the same time the estimation of a regression model for each group. However, covariates can not only be linked to the distribution parameters of each mixture component but also to the mixture weights as further discussed by \textcite{Wedel2002a}. For a brief overview of GLM based mixture regression models, see, e.g. \textcites{DeSarbo1994, Wedel2000, Gruen2008}. Mixture regression models following the idea of a fully distributional regression approach have been suggested in the setting of generalized additive models for location, scale and shape (GAMLSS, \textcite{Stasinopoulos2006}), where also nonlinear terms are allowed to affect the distribution parameters \parencite{Stasinopoulos2017}.

In the following, the notion of a mixture regression model is more mathematically introduced  in alignment with \textcite{Kneib2023}. The goal is to model the response variable $Y\in \R$ conditional on covariates $\bm{X}=\bm{x}\in \R^p$ for which a set of $N$ independent realizations $\mng{(y^{(i)}, \bm{x}^{(i)})}_{i=1,\ldots, N}$ is available. The key assumption is, that the PDF of the conditional distribution $\mathcal{D}(Y\,\vert\,\bm{X}=\bm{x})$ is given by
\begin{align}
f(y\,\vert\,\bm{x}):=\sum\limits_{k=1}^{K}\omega_k(\bm{x})f_k(y\,\vert\, \bm{\theta}_k(\bm{x})), \label{eq: MRM}
\end{align}
which is a mixture of parametric PDFs $f_k$ with corresponding distribution parameters $\bm{\theta}_k(\bm{x})$ influenced by a subset of covariates of $\bm{x}\in \R^p$. The number of mixture components $K$ is assumed to be known for the rest of the paper and the selected parametric distribution family $\mathcal{D}_k(\bm{\theta}_k)$ can differ among each mixture component $f_k$. Furthermore, the mixture weights $\omega_k(\bm{x})\in[0,1]$ depend on a subset of covariates of $\bm{x}\in \R^p$ as well, and need to fulfill the condition $\sum\limits_{k=1}^{K}\omega_k(\bm{x})=1$. Hence the vector of all $J:=K+D$ parameters $\bm{\psi}:=(\bm{\omega}, \bm{\theta})\in \R^{K+D}$ for the model in Equation \eqref{eq: MRM} consists of $K$ mixture weights 
\linebreak$\bm{\omega}:=(\omega_1,\ldots,\omega_K)$ and $D:=\sum\limits_{k=1}^{K}D_k$ distribution parameters $\bm{\theta}:=(\bm{\theta}_1,\ldots, \bm{\theta}_K)$. Throughout the paper, the covariates are linked to the mixture weights and to the distribution parameters of each mixture component in terms of GLMs, although more sophisticated models, such as  generalized additive models (GAMs, \cite{Hastie1986, Green1993}), could be employed. More precisely, each parameter 
\begin{align}
\psi_j(\bm{x}):=g_j(\eta_j(\bm{x})):=g_j(\alpha_{0,j}+\alpha_{1,j}x_{1,j}+\ldots+\alpha_{p_j,j}x_{p_j,j}),\quad j=1,\ldots, J,
\end{align}
depends on a linear predictor $\eta_j$ with coefficients $\bm{\alpha}_j:=(\alpha_{0,j}, \alpha_{1,j},\ldots,\alpha_{p_j,j})\in \R^{p_j+1}$, which is linked via a monotonic and differentiable link function $g_j$ to the corresponding parameter $\psi_j$, to ensure the correct parameter domain. 
For the remainder of the manuscript, the softmax function \parencite{Bishop1994, Stasinopoulos2017} is assumed to relate the first $K$ linear predictors $\eta_j$ to their corresponding mixture weights
\begin{align} \omega_j(\bm{x}):=\frac{\exp(\eta_j(\bm{x}))}{\sum\limits_{k=1}^{K}\exp(\eta_k(\bm{x}))},\quad j=1,\ldots,K,
\end{align}
which guarantees the previously mentioned requirements for mixture weights. 
Note, that a linear predictor $\eta_j$ might have only an intercept or depends on possibly different subsets of $1\leq p_j\leq p$ covariates $x_{1,j},\ldots, x_{p_j,j}$ of $\bm{x}\in \R^p$. 

\subsection{Gradient-boosting framework}
\label{sec: Gradient-boosting framework}

The boosting framework has initially been proposed for classification tasks by \textcite{Freund1996} and has been adapted to gradient-based regression settings most notably by \textcites{Friedman2001, Buehlmann2003, Buehlmann2007, Mayr2012}. Recently, \textcite{Hepp2023} have embedded mixture regression models in a gradient-boosting setting. They utilize the cyclic gradient-boosting  algorithm \parencite{Mayr2012} which iteratively updates for every linear predictor, the coefficient associated with the covariate most enhancing the current fit. In contrast, \textcite{Messner2017} have proposed the later termed non-cyclic gradient-boosting algorithm which updates a single coefficient across all linear predictors in each iteration, specifically the one linked to the covariate reducing the loss the most. Consequently, if the loss optimization is halted by a stopping criterion prior to achieving convergence, only the most relevant covariates obtain a non-zero coefficient, thus are selected, whereas less important covariates have a zero coefficient and are therefore disregarded. As a result, the estimation algorithm performs an intrinsic variable selection, is therefore robust against previously mentioned multicollinearity issues as well as overfitting and leads in total to more accurate forecasts \parencite{Mayr2018}. \textcite{Thomas2017} have illustrated in a simulation study that the non-cyclic variant can surpass the cyclic version regarding convergence speed and computational efficiency, thus making it an appealing choice for gradient-boosting applications. Therefore, the non-cyclic gradient-boosting algorithm for mixture regression models is summarized in Algorithm \ref{alg: boosting} and explained in more detail, subsequently. 

Similar to \textcite{Messner2017}, it is assumed that the realizations $\mng{(y^{(i)}, \bm{x}^{(i)})}_{i=1,\ldots, N}$ of the response variable and the covariates are standardized and have therefore zero mean and unit variance, which slightly simplifies the gradient-boosting estimation. Additionally, this allows to directly compare the covariates with respect to their importance. For the \textit{loss function} $\ell$ of the conditional distribution $\mathcal{D}(Y\,\vert\,\bm{X}=\bm{x})$, either the logarithmic score \parencite{Good1952} or the continuous ranked probability score \parencite{Matheson1976} is considered, see section \ref{sec: Forecast evaluation}.

\algnewcommand\algorithmicinit{\textbf{Initialization}}
\algnewcommand\init{\item[\algorithmicinit]}

\algnewcommand\algorithmicboost{\textbf{Boosting}}
\algnewcommand\boost{\item[\algorithmicboost]}

\algnewcommand\algorithmicfinal{\textbf{Finalization}}
\algnewcommand\final{\item[\algorithmicfinal]}

\begin{algorithm}[h!]
\caption{Non-cyclic gradient-boosting for mixture regression models}
\begin{algorithmic}

\init  Initialize
\begin{align*}
\bm{\alpha}_{j}^{[0]}:=(\alpha_{1,j}^{[0]},\ldots,\alpha_{p_j, j}^{[0]}):=(0,\ldots,0),\quad j=1,\ldots,J.
\end{align*}

\boost For each iteration $m = 1,\ldots, m_{\text{stop}}$:
\State	1. Calculate the negative gradient of the loss function $\ell$ with respect to each linear predictor $\eta_{j}$:

$$\left(g_{j}^{(i)}\right)_{i=1,\ldots,N}:=\left(-\frac{\partial}{\partial \eta_j}\ell\big(y^{(i)}; \eta_{1}^{[m-1]}(\bm{x}^{(i)}),\ldots, \eta_{J}^{[m-1]}(\bm{x}^{(i)})\big)\right)_{i=1,\ldots, N},\quad j=1,\ldots,J.$$

\State 2. Fit for all $p_j$ covariates $x_{1,j},\ldots,x_{p_j,j}$ of each linear predictor $\eta_{j}$ a separate linear regression model without intercept in the sense of ordinary least squares to the negative gradient:
$$g_{j}^{(i)}=\widehat{\rho}_{k,j} x_{k,j}^{(i)},\quad i=1,\ldots,N,$$
with slope $\displaystyle \widehat{\rho}_{k,j}=\frac{1}{N}\sum\limits_{i=1}^{N}x_{k,j}^{(i)}g_{j}^{(i)}$ for each $k=1,\ldots, p_j$, $j=1,\ldots,J$.

\State 3. Select for each linear predictor $\eta_{j}$ the index $k_j$ which minimizes the residual sum of squares criterion:
$$k_j:=\argmin\limits_{k=1,\ldots,p_j} \; \sum\limits_{i=1}^{N}\left(g_{j}^{(i)}-\widehat{\rho}_{k,j} x_{k,j}^{(i)}\right)^2\quad \Leftrightarrow\quad k_j:=\argmax\limits_{k=1,\ldots,p_j} \; |\widehat{\rho}_{k,j}|,\quad j=1,\ldots,J.$$

\State 4. Calculate the potential loss by updating only a single linear predictor $\eta_j$:
{\small
$$\Delta \ell_j:=\sum\limits_{i=1}^{N} \ell\big(y^{(i)};\, \eta_{1}^{[m-1]}(\bm{x}^{(i)})+\mathds{1}_{\{j=1\}}\nu\widehat{\rho}_{k_1,1}x_{k_1,1}^{(i)},\ldots, \eta_{J}^{[m-1]}(\bm{x}^{(i)})+\mathds{1}_{\{j=J\}}\nu\widehat{\rho}_{k_{J} ,J}x_{k_{J},J}^{(i)})\big),$$
}
\State using step length $\nu\in (0,1]$ for each $j=1,\ldots,J$, where $\mathds{1}$ denotes the indicator function.

\State 5. Select the index $j^\ast$ for which the updated linear predictor $\eta_{j^\ast}$ yields the lowest potential loss:
$$j^\ast:=\argmin\limits_{j=1,\ldots,J}\Delta\ell_j.$$

\State 6. Really update only the coefficient of the covariate improving the current fit the most:
\begin{align*}
\alpha_{k_{j^\ast},j^\ast}^{[m]}&:=\alpha_{k_{j^\ast},j^\ast}^{[m-1]}+\nu\widehat{\rho}_{k_{j^\ast},j^\ast},\quad \alpha_{k,j^\ast}^{[m]}:=\alpha_{k,j^\ast}^{[m-1]},\quad k\in \mng{1,\ldots,p_{j^\ast}}\backslash\mng{k_{j^\ast}},\\
\bm{\alpha}_j^{[m]}&:=\bm{\alpha}_j^{[m-1]},\quad j\in \{1,\ldots,J\}\backslash\{j^\ast\},
\end{align*}
\State using step length $\nu\in (0,1]$.

\final Set 
\begin{align*}
\bm{\alpha}_j:=\bm{\alpha}_j^{[m_{\mathrm{stop}}]},\quad j = 1,\ldots, J. 
\end{align*}
\end{algorithmic}
\label{alg: boosting}
\end{algorithm} 

The algorithm starts with initializing every coefficient across all linear predictors with zero. During each boosting iteration, the negative gradient of the loss function with respect to each linear predictor is computed. Subsequently, individual linear regression models are fitted to the negative gradient $-$ in terms of the ordinary least squares method $-$ for all covariates and each linear predictor. In the next step, only one coefficient of each linear predictor is selected for an subsequent update according to the minimum residual sum of squares criterion employing the previously fitted linear regression models without intercept. Note, that due to the assumption of standardized covariates, the previously mentioned coefficient selection can be simplified by searching the covariate which has the highest absolute covariance with the corresponding negative gradient of the loss function. By construction, the covariance is equivalent to the slope of the linear regression model. In the next step, the product of the selected covariate and the corresponding slope of the fitted linear regression model are further scaled by a fixed \textit{step length} $\nu\in(0,1]$. According to \textcite{Buehlmann2007}, the step length is a tuning parameter of minor importance, as long as a small value, e.g. $\nu=0.1$, is chosen. During this step, the potential loss is obtained by adding the previously calculated product to the respective linear predictor, while the remaining linear predictors steam from the previous boosting iteration. Eventually, only the coefficient corresponding to the linear predictor reducing the loss the most is updated, while the coefficients of all remaining linear predictors are taken from the previous boosting iteration.
This whole process is repeated until the \textit{maximum number of boosting iterations}, denoted as $m_{\mathrm{stop}}$, is reached. In order to avoid overfitting, the \textit{optimal number of boosting iterations} $m_{\mathrm{opt}}$ needs to be found after this initial gradient-boosting. Similar to \textcite{Messner2017}, a $K$-fold cross validation is employed to find $m_{\mathrm{stop}}$. The training data is split into $K$ random folds where for all possible fold combinations, $K-1$ folds are used for training and the remaining fold is employed for validation with respect to the loss function. This procedure is carried out for boosting iterations 1 to $m_{\mathrm{stop}}$. The iteration providing the lowest total loss over all validation folds is selected as $m_{\mathrm{opt}}$.

\section{Data}
\label{sec: Data}

In this section, the observation and forecast data as well as its usage for the case study are presented. Furthermore, the related notation is given. 

\begin{figure}[h!]
	\begin{center}
		\includegraphics[scale = 0.5]{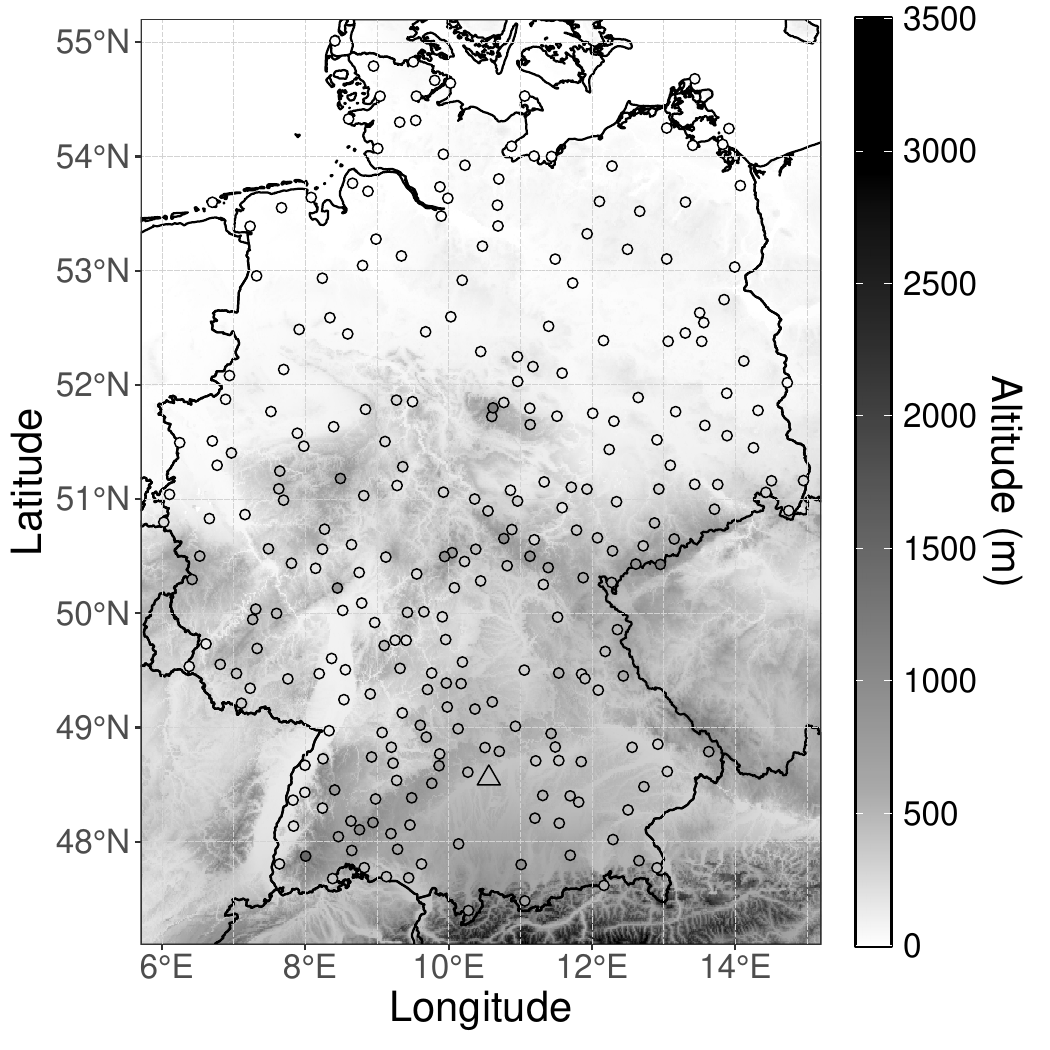}
	\end{center}	
	\caption{Location of 280 observation stations in Germany displayed on the topography from the Shuttle Radar Topography Mission at 90\,\si{\meter} resolution \parencite{Reuter2007}. Station Dillingen/Donau-Fristingen in the south is marked by $\triangle$.}
	\label{fig: stations}
\end{figure}

\paragraph{Observation and forecast data.}
The observation data contains 2\,\si{m} surface temperature measurements at 280 SYNOP stations in Germany as shown in Figure \ref{fig: stations}. This data is provided by the German weather service \parencite{DWD2018} and covers the period from January 2, 2015 to December 31, 2020 (2191 days). Moreover, less than 1\% of observations are missing at each station, which are not further pre-processed.  

The forecast data is supplied by the \textcite{ECMWF2021}, containing 50 perturbed ensemble forecasts including one control forecast for eight different weather variables, see Table \ref{tab: pred_var}. These forecasts are initialized at 1200 UTC and are valid for a lead time of 24\,\si{\hour}. As these forecasts are discretized on a grid with resolution of 0.25$^\circ$ $\times$ 0.25$^\circ$ ($\approx$ 28\,\si{km} squared), they are bilinearly interpolated from the closest four surrounding grid points to the stations. 

\begin{table}[h!]
\begin{center}
\resizebox{15cm}{!}{
\begin{tabular}{c c c c c c}
\toprule
\multirow{2}{*}{Abbreviation ($w$)} & \multirow{2}{*}{Description} & \multirow{2}{*}{Unit} & \multicolumn{3}{c}{ Transformation $h$ for } \\
\cline{4-6} 
 & & & $X_{w}^{\mathrm{mean}}$ & $X_{w}^{\mathrm{ctrl}}$ & $X_{w}^{\mathrm{sd}}$ \\
\hline
t2m & 2\,\si{\meter} surface temperature & \si{\degreeCelsius} & $\mathrm{id}(\cdot)$ & $\mathrm{id}(\cdot)$ & $\log(\cdot)$ \\
pr & surface pressure & \si{\hecto\pascal} & $\mathrm{id}(\cdot)$ & $\mathrm{id}(\cdot)$ & $\log(\cdot)$\\
u10m & 10\,\si{\meter} surface $u$-wind speed component & \si{\meter\per\second} & $\mathrm{id}(\cdot)$ & $\mathrm{id}(\cdot)$ & $\log(\cdot)$\\
v10m & 10\,\si{\meter} surface $v$-wind speed component & \si{\meter\per\second} & $\mathrm{id}(\cdot)$ & $\mathrm{id}(\cdot)$ & $\log(\cdot)$\\
sh & specific humidity & \si{\kilogram\per\kilogram} & $\log(\cdot)$ & $\log(\cdot)$ & $\log(\cdot)$\\
tcc & total cloud cover & $-$ & $\mathrm{logit}(\cdot)$ & $\mathrm{logit}(\cdot)$ & $\mathrm{logit}(\cdot/2)$\\
ws10m & 10\,\si{\meter} surface wind speed & \si{\meter\per\second}  & $\log(\cdot)$ & $\log(\cdot)$ & $\log(\cdot)$\\
wg10m & 10\,\si{\meter} surface wind gust & \si{\meter\per\second}  & $\log(\cdot)$ & $\log(\cdot)$ & $\log(\cdot)$\\
\bottomrule
\end{tabular}
}
\end{center}
\caption{Available predictor variables.}
\label{tab: pred_var}
\end{table}

\paragraph{Training, validation and testing data.}
The data between January 2, 2015 to December 31, 2019 (1825 days) serves as static training data for all methods. Additionally, all methods estimate local ensemble postprocessing models, i.e. for each station a separate model is fitted. Note, that methods, estimated via gradient-boosting require parameter tuning via $K$-fold cross validation (CV). Consequently, parts of the training data are used as validation data for these methods. Finally, all methods are compared on the testing data between January 1, 2020 to December 31, 2020 (366 days).

\paragraph{Notation.}
In the following, the response variable 2\,\si{m} surface temperature is denoted by $Y$. Furthermore, $X_{w}^{p, 1}, \ldots, X_{w}^{p, 50}$ as well as $X_{w}^{\mathrm{ctrl}}$ represent the 50 perturbed ensemble forecasts and the control forecast for any weather variable $w$ from the set $W:=\{\mathrm{t2m}, \mathrm{pr}, \mathrm{u10m}, \mathrm{v10m}, \mathrm{sh}, \mathrm{tcc}, \mathrm{ws10m}, \mathrm{wg10m}\}$ of all available weather variables. The 50 perturbed ensemble forecasts as well as the control forecast form two separate exchangeable groups. Therefore, the $n=50$ perturbed ensemble forecasts can be reduced to the empirical ensemble mean and its empirical ensemble standard deviation 
\begin{align}
    X_{w}^{\mathrm{mean}}:=\frac{1}{n}\sum\limits_{i=1}^{n}X_{w}^{p,i},\quad X_{w}^{\mathrm{sd}}:=\sqrt{\frac{1}{n-1}\sum\limits_{i=1}^{n}\left(X_{w}^{p,i}-X_{w}^{\mathrm{mean}}\right)^2},
\end{align}
for any weather variable  $w\in W$. Eventually, the realizations of all variables are denoted by lower-case letters.

\section{Ensemble postprocessing methods}
\label{sec: Ensemble postprocessing methods}

Given that all subsequently presented ensemble postprocessing models utilize standardized anomalies, this section begins with an introduction to their calculation. Afterwards, the state-of-the-art ensemble postprocessing methods SAMOS and SAMOS-GB are described. Furthermore, the framework for the mixture regression models MIXMOS, MIXSAMOS as well as its gradient-boosted extension MIXSAMOS-GB are explained.  Note, that all presented postprocessing models are completely optimized either via the logarithmic score or via the continuous ranked probability score using the \texttt{R}-packages \texttt{crch} \parencite{Messner2013a} and \texttt{mixnhreg} \parencite{Jobst2024c}. Additional information about specified hyperparameters for all methods can be found in Appendix \ref{app: Hyperparameters}.

\subsection{Standardized anomalies}
\begin{figure}[h!]
	\begin{center}
		\includegraphics[scale = 0.6]{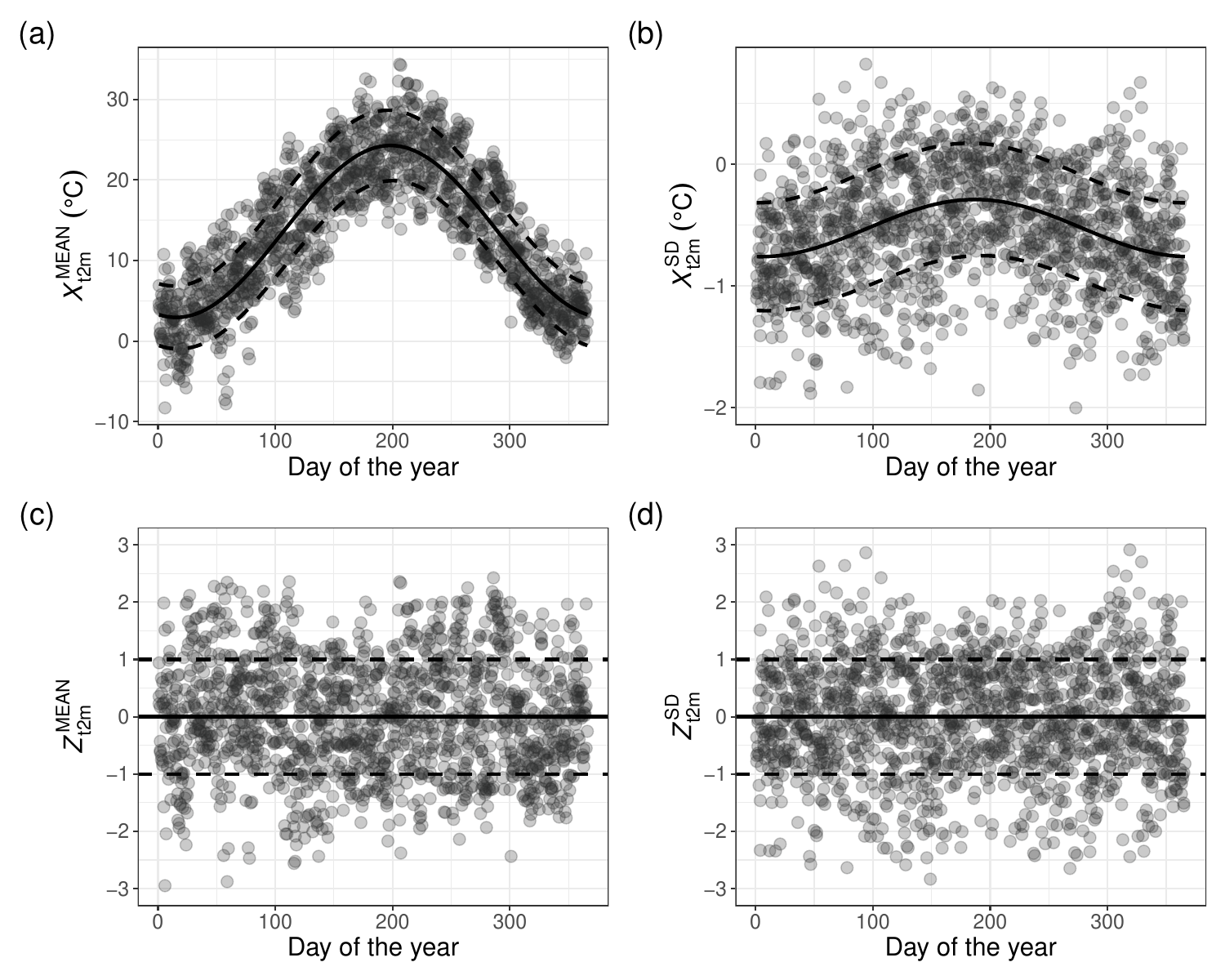}
	\end{center}	
	\caption{Upper panels (a), (b): ensemble mean $X_{\mathrm{t2m}}^{\mathrm{MEAN}}$ and standard deviation $X_{\mathrm{t2m}}^{\mathrm{SD}}$ of 2\,\si{\metre} surface temperature for Dillingen/Donau-Fristingen (gray points), corresponding estimated climatological mean (solid line) and climatological mean $\pm$ climatological standard deviation (dashed lines). Lower panels (c), (d): calculated standardized anomalies $Z_{\mathrm{t2m}}^{\mathrm{MEAN}}$ and $Z_{\mathrm{t2m}}^{\mathrm{SD}}$ (gray points)  of $X_{\mathrm{t2m}}^{\mathrm{MEAN}}$ and $X_{\mathrm{t2m}}^{\mathrm{SD}}$, respectively.}
	\label{fig: anomalies}
\end{figure}

With the growing abundance of accessible data for ensemble postprocessing, including benchmark studies like that of \textcite{Demaeyer2023}, there is currently a transition from employing rolling training windows towards utilizing static training, resulting in a more parsimonious model estimation. A longer static training data can considerably improve the predictive model performance as shown by \textcite{Lang2020} and has therefore found its application in operational settings \parencite{Hess2020}, too. Nevertheless, whereas models utilizing rolling training windows account for temporal effects by their design, those relying on static training data must explicitly incorporate these effects into their framework. To remove the seasonal effects inherent in the training data and to leverage a longer static training data, \textcite{Dabernig2017} suggested to employ standardized anomalies of the considered variables instead of the original data. In order to calculate  standardized anomalies, seasonally varying climatological means and standard deviations are derived for the response and all covariates via a non-homogeneous linear regression model in a first step. To facilitate the estimation procedure, assume that all considered variables follow a normal distribution, which additionally requires to apply a variable transformation $h$ (see Table \ref{tab: pred_var}) on the covariates $X_{w}^{\mathrm{mean}}, X_{w}^{\mathrm{ctrl}}, X_{w}^{\mathrm{sd}}$ resulting in
\begin{align}
X_{w}^{\mathrm{MEAN}}:=h\left(X_{w}^{\mathrm{mean}}\right),\quad X_{w}^{\mathrm{CTRL}}:=h\left(X_{w}^{\mathrm{ctrl}}\right),\quad X_{w}^{\mathrm{SD}}:=h\left(X_{w}^{\mathrm{sd}}\right),  
\end{align}
for all weather variables $w\in W$. Consequently, for any variable $\xi \in \{Y, X_{w}^{\mathrm{MEAN}}, X_{w}^{\mathrm{CTRL}}, X_{w}^{\mathrm{SD}}\}_{w\in W}$ the following non-homogenous regression model
\begin{gather}
\xi \sim \mathcal{N}(\mu_\xi,\sigma_\xi^2),\\
\begin{alignedat}{1}
g_1^{-1}(\mu_\xi(\mathrm{doy}))&=\eta_1(\mathrm{doy})=\underbrace{\alpha_{0,1}+\alpha_{1,1}\sin\left(2\pi\frac{\mathrm{doy}}{365.25}\right)+\alpha_{2,1}\cos\left(2\pi\frac{\mathrm{doy}}{365.25}\right)}_{\text{seasonal varying intercept}},\\  
g_2^{-1}(\sigma_\xi(\mathrm{doy}))&=\eta_2(\mathrm{doy})=\underbrace{\alpha_{0,2}+\alpha_{1,2}\sin\left(2\pi\frac{\mathrm{doy}}{365.25}\right)+\alpha_{2,2}\cos\left(2\pi\frac{\mathrm{doy}}{365.25}\right)}_{\text{seasonal varying intercept}}, \label{eq: climatology_parameter}
\end{alignedat}
\end{gather}
with inverse links functions $g_1^{-1}=\mathrm{id}$, $g_2^{-1}=\log$ is estimated, where $\mathrm{doy}\in \{1,2,\ldots, 366\}$ stands for the day of the year and $\alpha_{i,j}\in \R$ denote the regression coefficients. The intercept of the location and scale parameter in Equation \eqref{eq: climatology_parameter} are both modeled by one sine and one cosine base function depending on the day of the year to capture seasonal effects \parencite{Messner2017}. The model estimation is performed via the Broyden-Fletcher-Goldfarb-Shanno (BFGS) algorithm in the training data.  In the final step, the standardized anomalies are calculated for the response 
\begin{align}
    Z:=\frac{Y-\mu_Y}{\sigma_Y},
\end{align}
and the covariates
\begin{align}
Z_{w}^{\mathrm{MEAN}}:=\frac{X_{w}^{\mathrm{MEAN}}-\mu_{X_{w}^{\mathrm{MEAN}}}}{\sigma_{X_{w}^{\mathrm{MEAN}}}},\, Z_{w}^{\mathrm{CTRL}}:=\frac{X_{w}^{\mathrm{CTRL}}-\mu_{X_{w}^{\mathrm{CTRL}}}}{\sigma_{X_{w}^{\mathrm{CTRL}}}},\, Z_{w}^{\mathrm{SD}}:=\frac{X_{w}^{\mathrm{SD}}-\mu_{X_{w}^{\mathrm{SD}}}}{\sigma_{X_{w}^{\mathrm{SD}}}},
\label{eq: standardized covariates}
\end{align}
of all weather variables $w\in W$. Consequently, the standardized response and standardized covariates are standard normally distributed. Figure \ref{fig: anomalies} exemplary shows that after subtracting the climatological mean and dividing by the climatological standard deviation, the calculated standardized anomalies of $X_{\mathrm{t2m}}^{\mathrm{MEAN}}$ and $X_{\mathrm{t2m}}^{\mathrm{SD}}$ contain no pronounced seasonal effects in the training data anymore. 

\subsection{State-of-the-art ensemble postprocessing methods}

In the following, the considered benchmark methods are presented. 

\paragraph{Standardized anomaly model output statistics.}
Some years after the introduction of the \textit{standardized anomaly model output statistic} (SAMOS, \cite{Dabernig2017}), this method is run meanwhile operationally by GeoSphere Austria. In order to obtain probabilistic SAMOS forecasts, a heteroscedastic linear regression model is estimated on the standardized anomalies of the response and covariates in a first step. In this setting, the standardized response follows a normal distribution and the distribution parameters are specified as follows:
\begin{gather}
Z \sim \mathcal{N}(\mu_Z(\bm{x}),\sigma_Z(\bm{x})^2), \label{eq: SAMOS-Z}\\
\begin{alignedat}{1}
g_1^{-1}(\mu_Z(\bm{x}))=\eta_1(\bm{x})&=\alpha_{0,1}+\alpha_{1,1}z_{1,1}+\alpha_{2,1}z_{2,1}=\alpha_{0,1}+\alpha_{1,1}z_{\mathrm{t2m}}^{\mathrm{MEAN}}+\alpha_{2,1}z_{\mathrm{t2m}}^{\mathrm{CTRL}}\\
&=\alpha_{0,1}+\alpha_{1,1}\frac{x_{\mathrm{t2m}}^{\mathrm{MEAN}}-\mu_{X_{\mathrm{t2m}}^{\mathrm{MEAN}}}}{\sigma_{X_{\mathrm{t2m}}^{\mathrm{MEAN}}}}+\alpha_{2,1}\frac{x_{\mathrm{t2m}}^{\mathrm{CTRL}}-\mu_{X_{\mathrm{t2m}}^{\mathrm{CTRL}}}}{\sigma_{X_{\mathrm{t2m}}^{\mathrm{CTRL}}}},\\
g_2^{-1}(\sigma_Z(\bm{x}))=\eta_2(\bm{x})&=\alpha_{0,2} + \alpha_{1,2}z_{1,2}=\alpha_{0,2} + \alpha_{1,2}z_{\mathrm{t2m}}^{\mathrm{SD}}\\
&=\alpha_{0,2} + \alpha_{1,2}\frac{x_{\mathrm{t2m}}^{\mathrm{SD}}-\mu_{X_{\mathrm{t2m}}^{\mathrm{SD}}}}{\sigma_{X_{\mathrm{t2m}}^{\mathrm{SD}}}},\label{eq: SAMOS-Z-parameter}
\end{alignedat}
\end{gather}
with regression coefficients $\alpha_{i,j}\in \R$ of the standardized anomalies $z_{i,j}$ and inverse link functions $g_1^{-1}=\mathrm{id}$, $g_2^{-1}=\log$. The model estimation is carried out on the training data utilizing the BFGS algorithm similar to \textcite{Gneiting2005}. 

As the predictive distribution in Equation \eqref{eq: SAMOS-Z} acts on the standardized scale, its distribution parameters need to be destandardized in a second step. This can be accomplished by rescaling the distribution parameters on the standardized scale by the distribution parameters of the climatology of the response. Consequently, the predictive distribution for the observed scale is given by
\begin{gather}
Y \sim \mathcal{N}(\mu(\bm{x}),\sigma(\bm{x})^2),\label{eq: SAMOS model}\\
\begin{alignedat}{1}
\mu(\bm{x})=\mu_Z(\bm{x})\cdot \sigma_Y(\mathrm{doy}) + \mu_Y(\mathrm{doy}), \quad \sigma(\bm{x})=\sigma_Z(\bm{x})\cdot \sigma_Y(\mathrm{doy}), \label{eq: SAMOS-parameter}\\
\end{alignedat}
\end{gather}
where $\mu_Z(\bm{x}), \sigma_Z(\bm{x})$ denote the mean and standard deviation of the EMOS like model in \eqref{eq: SAMOS-Z-parameter} and $\mu_Y(\mathrm{doy}), \sigma_Y(\mathrm{doy})$ stand for the climatological mean and standard deviation of the response in Equation \eqref{eq: climatology_parameter}. Note, that by replacing the mean $\mu_Z(\bm{x})$ in Equation \eqref{eq: SAMOS-parameter} by its expression in Equation \eqref{eq: SAMOS-Z-parameter}, one can prove that every member within a single exchangeable group is assigned an identical regression coefficient. Consequently, the assumption of exchangeable groups from the EMOS model is preserved by the SAMOS model, which has neither been explicitly discussed, nor investigated, yet.

\paragraph{Gradient-boosted standardized anomaly model output statistics.}

The SAMOS model only includes the standardized anomalies of the ensemble mean, of the control forecast as well as of the ensemble standard deviation. However $8\cdot 3 = 24$ potential covariates are available, see Table \ref{tab: pred_var}. Selecting the most relevant requires knowledge and can be laborious for many stations and/or lead times. Therefore, \textcite{Messner2017} suggest to estimate the \textit{SAMOS model via gradient-boosting}, called SAMOS-GB model. Similar to SAMOS, SAMOS-GB assumes a normal distribution for the standardized response
\begin{gather}
Z \sim \mathcal{N}(\mu_Z(\bm{x}),\sigma_Z(\bm{x})^2), \label{eq: SAMOS-GB-Z}\\
\begin{alignedat}{1}
g_1^{-1}(\mu_Z(\bm{x})&=\eta_1(\bm{x})=\alpha_{0,1}+\alpha_{1,1}z_{1,1}+\ldots+\alpha_{p_1,1}z_{p_1,1},\\  
g_2^{-1}(\sigma_Z(\bm{x}))&=\eta_2(\bm{x})=\alpha_{0,2}+\alpha_{1,2}z_{1,2}+\ldots+\alpha_{p_2,2}z_{p_2,2}.\label{eq: SAMOS-GB-Z-parameter}
\end{alignedat}
\end{gather}
The $p_1=p_2=24$ calculated standardized anomalies $z_{i,j}$ of all weather variables in Equation \eqref{eq: standardized covariates} are employed as covariates for both linear predictors. The latter are connected through the inverse link functions $g_1^{-1}=\mathrm{id}$, $g_2^{-1}=\log$ to the distribution parameters in Equation \eqref{eq: SAMOS-GB-Z-parameter}, where $\alpha_{i,j}\in \R$ denote its coefficients. The model specified in Equations \eqref{eq: SAMOS-GB-Z}, \eqref{eq: SAMOS-GB-Z-parameter} is initially estimated via the non-cyclic gradient-boosting algorithm on the training data. In order to avoid overfitting, $m_{\mathrm{opt}}$ is afterwards determined via CV. Then, the predictive distribution of SAMOS-GB can be recovered for the observed scale analogously to SAMOS via Equations \eqref{eq: SAMOS model}, \eqref{eq: SAMOS-parameter}. Note, that SAMOS-GB supports the notion of exchangeable groups, too.

\subsection{Mixture regression models for ensemble postprocessing}

The novel mixture regression models for ensemble postprocessing are subsequently introduced. 

\paragraph{Mixture of model output statistics.}
 
The baseline model for all upcoming models is given by the \textit{mixture of model output statistics} (MIXMOS). A slightly simpler version thereof has been proposed by \textcite{Taillardat2021}, who linked the covariates only to the parameters of the mixture components, but not to the mixture weights. The PDF of the MIXMOS conditional distribution $\mathcal{D}(Y\,\vert\,\bm{X}=\bm{x})$ is defined as
\begin{align}
f(y\,\vert\,\bm{x})=\sum\limits_{k=1}^{K}\omega_k(\bm{x}_k)f_k(y\,\vert\, \bm{\theta}_k(\bm{x}_k)), \label{eq: MIXMOS}
\end{align}
where $\bm{x}_k$ is a subset vector of all covariates $\bm{x}\in \R^p$  only containing the ensemble forecasts (or its summary statistics) of the $k$-th exchangeable group. Besides this restriction, the mixture regression model is defined as introduced in section \ref{sec: General model definition}. In contrast to BMA, MIXMOS has less mixture components $f_k$ and consequently also less constraints for all parameters providing a somehow simpler and more parsimonious model. At the same time, the distribution family $\mathcal{D}_k(\bm{\theta}_k)$ of each mixture component $f_k$ in MIXMOS can be individually specified in contrast to BMA, making MIXMOS a flexible distributional postprocesssing approach. 

For example, there exist $K=2$ exchangeable groups in the setting of the case study, where one group consists of the 50 perturbed temperature forecasts, which can be reduced to its ensemble mean and  ensemble standard deviation, denoted as $\bm{x}_1:=(x_{\mathrm{t2m}}^{\mathrm{MEAN}}, x_{\mathrm{t2m}}^{\mathrm{SD}})$. The other group, denoted as $\bm{x}_2:=(x_{\mathrm{t2m}}^{\mathrm{CTRL}})$, is only based on the deterministic temperature control forecast. Under the common assumption $f_1\sim \mathcal{N}(\mu_1(\bm{x}_1),\sigma_1(\bm{x}_1)^2), f_2\sim \mathcal{N}(\mu_2(\bm{x}_2),\sigma_2(\bm{x}_2)^2)$ for 2\,\si{\meter} surface temperature observations, the conditional PDF of MIXMOS is given by
\begin{align}
f(y\,\vert\,\bm{x})=\omega_1(\bm{x}_1)f_1(y\,\vert\, \mu_1(\bm{x}_1), \sigma_1(\bm{x}_1)^2)+\omega_2(\bm{x}_2)f_2(y\,\vert\, \mu_2(\bm{x}_2), \sigma_2(\bm{x}_2)^2).   \label{eq: MIXMOS_example}
\end{align}
Subsets of the vectors $\bm{x}_1, \bm{x}_2$ can be linked as covariates to the distribution parameters. However, as all considered variables show a seasonal behavior, this effect should first be taken into account in order use a longer static training data and to obtain a reasonable predictive model. Therefore, the baseline MIXMOS model is not sufficient in its current form and subsequently extended employing standardized anomalies.

\paragraph{Mixture of standardized anomaly model output statistics.}

Similar to SAMOS, the standardized anomalies of the response and of the covariates are employed for the \textit{mixture of standardized anomaly model output statistics} (MIXSAMOS). Consequently, the mixture regression model for the $K=2$ exchangeable groups $\bm{x}_1=(x_{\mathrm{t2m}}^{\mathrm{MEAN}}, x_{\mathrm{t2m}}^{\mathrm{SD}})$, $\bm{x}_2=(x_{\mathrm{t2m}}^{\mathrm{CTRL}})$ is estimated on the standardized scale instead on the original scale with given conditional PDF
\begin{align}
f(z\,\vert\,\bm{x})=\omega_1(\bm{x}_1)f_1(z\,\vert\, \mu_{Z_1}(\bm{x}_1), \sigma_{Z_1}(\bm{x}_1)^2)+\omega_2(\bm{x}_2)f_2(z\,\vert\, \mu_{Z_2}(\bm{x}_2), \sigma_{Z_2}(\bm{x}_2)^2),  \label{eq: MIXMOS_example2}
\end{align}
where $f_1\sim \mathcal{N}(\mu_{Z_1}(\bm{x}_1),\sigma_{Z_1}(\bm{x}_1)^2), f_2\sim \mathcal{N}(\mu_{Z_2}(\bm{x}_2),\sigma_{Z_2}(\bm{x}_2)^2)$. Initial tests have shown, that MIXSAMOS with non-constant mixture weights results into sharper forecasts, than those corresponding to MIXSAMOS with constant mixture weights. Therefore, the linear predictors $\eta_1,\eta_2$ of the mixture weights $\omega_1,\omega_2$ are parametrized via 
\begin{align}
\eta_1(\bm{x}_1)&=\alpha_{0,1}+\alpha_{1,1}z_{1,1}=\alpha_{0,1}+\alpha_{1,1}z_{\mathrm{t2m}}^{\mathrm{MEAN}},\\
\eta_2(\bm{x}_2)&=\alpha_{0,2}+\alpha_{1,1}z_{1,2}=\alpha_{0,2}+\alpha_{1,2}z_{\mathrm{t2m}}^{\mathrm{CTRL}},
\end{align}
with softmax link function. In addition, the distribution parameters of the first mixture component $f_1$ are specified comparable to SAMOS via
\begin{align}
g_3^{-1}(\mu_{Z_1}(\bm{x}_1))=\eta_3(\bm{x}_1)&=\alpha_{0,3}+\alpha_{1,3}z_{1,3}=\alpha_{0,3}+\alpha_{1,3}z_{\mathrm{t2m}}^{\mathrm{MEAN}},\\
g_4^{-1}(\sigma_{Z_1}(\bm{x}_1))=\eta_4(\bm{x}_1)&=\alpha_{0,4}+\alpha_{1,4}z_{1,4}=\alpha_{0,4}+\alpha_{1,4}z_{\mathrm{t2m}}^{\mathrm{SD}},
\end{align}
and for the second mixture component $f_2$ via
\begin{align}
g_5^{-1}(\mu_{Z_2}(\bm{x}_1))=\eta_5(\bm{x}_2)&=\alpha_{0,5}+\alpha_{1,5}z_{1,5}=\alpha_{0,5}+\alpha_{1,5}z_{\mathrm{t2m}}^{\mathrm{CTRL}},\\
g_6^{-1}(\sigma_{Z_2}(\bm{x}_2))=\eta_6(\bm{x}_2)&=\alpha_{0,6},
\end{align}
with regression coefficients $\alpha_{i,j}\in \R$ and inverse link functions $g_3^{-1}=g_5^{-1}=\mathrm{id}$, $g_4^{-1}=g_6^{-1}=\log$. MIXSAMOS is estimated utilizing the BFGS algorithm on the training data. Afterwards, the predictive PDF for the observed scale in Equation \eqref{eq: MIXMOS_example} can be derived similar to SAMOS. In more detail, the distribution parameters of the $k$-th mixture component $f_k\sim \mathcal{N}(\mu_k(\bm{x}_k),\sigma_k(\bm{x}_k)^2)$ in Equation \eqref{eq: MIXMOS_example} are given by 
\begin{align}
    \mu_k(\bm{x}_k)=\mu_{Z_k}(\bm{x}_k)\cdot \sigma_Y(\mathrm{doy}) + \mu_Y(\mathrm{doy}),\quad \sigma_k(\bm{x}_k)=\sigma_{Z_k}(\bm{x}_k)\cdot \sigma_Y(\mathrm{doy}), \label{eq: MIXSAMOS-parameter}
\end{align}
for $k = 1,2$. Note, that the mixture weights remain unchanged between the standardized and observed scale and that the exchangeability assumption remains valid for each mixture component.

\paragraph{Gradient-boosted mixture of standardized anomaly model output statistics.}

For MIXSAMOS only the ensemble forecasts corresponding to 2\,\si{\meter} surface temperature are allowed as covariates. However, additional weather variables might enhance the performance of MIXSAMOS, if carefully selected, as well. Consequently, the \textit{gradient-boosted mixture of standardized anomaly model output statistics} (MIXSAMOS-GB) is specified as in Equation \eqref{eq: MIXMOS_example2}, but with the difference, that the standardized anomalies of all available weather variables from Table \ref{tab: pred_var} are allowed for the first group $\bm{x}_1:=(x_{w}^{\mathrm{MEAN}}, x_{w}^{\mathrm{SD}})_{w\in W}$ and for the second group $\bm{x}_2:=(x_{w}^{\mathrm{CTRL}})_{w\in W}$. Consequently, each of the three linear predictors $\eta_j$ corresponding to the $k$-th mixture component is specified via
\begin{align}
    \eta_j(\bm{x}_k)=\alpha_{0,j}+\alpha_{1,j}z_{1,j}+\ldots + \alpha_{p_k,j}z_{p_k,j},\quad k=1,2.
\end{align}
The regression coefficients of the $p_1=16$ standardized anomalies $z_{i,j}$ for the first group and of the $p_2=8$ standardized anomalies $z_{i,j}$ for the second group are denoted by $\alpha_{i,j}\in \R$. Additionally, the same link functions as for MIXSAMOS are employed. The specified mixture regression model is estimated via the non-cyclic gradient-boosting algorithm introduced in section \ref{sec: Gradient-boosting framework} and summarized in Algorithm \ref{alg: boosting}. Afterwards, $m_{\mathrm{opt}}$ is determined via CV. Eventually, the predictive distribution for the observed scale in Equation \eqref{eq: MIXMOS_example} can be obtained as for MIXSAMOS in Equation \eqref{eq: MIXSAMOS-parameter}.

\section{Forecast evaluation}
\label{sec: Forecast evaluation}

This section summarizes common evaluation methods for probabilistic weather forecasts. For a more detailed introduction, please refer for example to \textcite{Vannitsem2018}.

\paragraph{Proper scoring rules.}
For model estimation and verification, proper scoring rules can be used as loss functions in order ``to maximize the sharpness of the predictive distribution function subject to calibration'' \parencite{Gneiting2007}. Among the most commonly employed proper scoring rules is the \textit{logarithmic score} (LogS, \cite{Good1952}), which is defined as 
\begin{align}
\mathrm{LogS}(F,y):=-\log(f(y)),
    \label{eq: LogS1}
\end{align}
where $f$ denotes the PDF of the predictive distribution $F$ evaluated at the verifying observation $y$. Another proper scoring rule, which takes account of the whole predictive distribution $F$ in contrast to the LogS, is the \textit{continuous ranked probability score} (CRPS, \cite{Matheson1976}). This score is evaluated for the predictive distribution $F$ at the verifying observation $y$ via
\begin{align}
\text{CRPS}(F,y):=\int\limits_{-\infty}^{\infty}(F(z)-\mathds{1}\{z\geq y\})^2\, \mathrm{d}z. \label{eq: CRPS1}
\end{align}
If $F$ is the predictive distribution function of a mixture regression model for which each mixture component in Equation \eqref{eq: MRM} follows a normal distribution, i.e. $f_k\sim \mathcal{N}(\mu_k(\bm{x}), \sigma_k(\bm{x})^2)$ for $k = 1,\ldots, K$, then the LogS is obtained via
\begin{align}
    \mathrm{LogS}(F,y)=-\log\left(\sum\limits_{k=1}^{K}\omega_k(\bm{x})f_k(y\,\vert\, \mu_k(\bm{x}), \sigma_k(\bm{x})^2)\right).
    \label{eq: LogS2}
\end{align}
Note, that the LogS in \eqref{eq: LogS2} accumulated over multiple forecast cases is equivalent to the negative log-likelihood of $f$. \textcite{Grimit2006} have proven under the same distribution assumption for $F$, that the CRPS can be calculated by
\begin{gather}
    \begin{alignedat}{1}
    &\mathrm{CRPS}(F,y)=\sum\limits_{k=1}^{K}\omega_k(\bm{x})A(y-\mu_k(\bm{x}), \sigma_k(\bm{x})^2)\\
    &-\frac{1}{2}\sum\limits_{k=1}^{K}\sum\limits_{j=1}^{K}\omega_k(\bm{x})\omega_j(\bm{x})A(\mu_k(\bm{x})-\mu_j(\bm{x}), \sigma_k(\bm{x})^2+\sigma_j(\bm{x})^2),\label{eq: CRPS2}
    \end{alignedat}
\end{gather}
where 
\begin{align}
A(\mu(\bm{x}), \sigma(\bm{x})^2):=\mu(\bm{x})\left(2\Phi\left(\frac{\mu(\bm{x})}{\sigma(\bm{x})}\right)-1\right)+2\sigma(\bm{x}) \varphi\left(\frac{\mu(\bm{x})}{\sigma(\bm{x})}\right),
\end{align}
and $\Phi, \varphi$ denote the distribution function and PDF of the standard normal distribution, respectively. If $F$ is just normally distributed, then its LogS and CRPS are given by Equation \eqref{eq: LogS2} and \eqref{eq: CRPS2} for $K=1$, $\omega_1 = 1$ \parencite{Gneiting2005}. 

\paragraph{Calibration and sharpness.}
While proper scoring rules assess calibration and sharpness of a predictive distribution $F$ jointly, it is also possible to verify these properties separately employing a $(1-\alpha)\cdot 100\%$, $\alpha\in (0,1)$ \textit{central prediction interval} $[l_\alpha,u_\alpha]\subseteq\mathbb{R}$ \parencite{Gneiting2007}. The lower and upper interval limits are given by $l_\alpha:=F^{-1}(\frac{\alpha}{2})$, $u_\alpha:=F^{-1}(1-\frac{\alpha}{2})$, respectively, where $F^{-1}$ stands for the quantile function of $F$. The \textit{width} of $[l_\alpha,u_\alpha]$ is given by  $u_\alpha-l_\alpha$ and measures the sharpness of $F$. The calibration of $F$ is assessed by the \textit{coverage} which is defined as the proportion of validating observations falling into $[l_\alpha,u_\alpha]$. In order to allow a direct comparison of a predictive distribution $F$ with a $m$-member forecast ensemble, a $\frac{m-1}{m+1}\cdot100\%$ central prediction interval is selected to match the ensembles nominal coverage.

\begin{figure}[h!]
	\begin{center}
		\includegraphics[scale = 0.5]{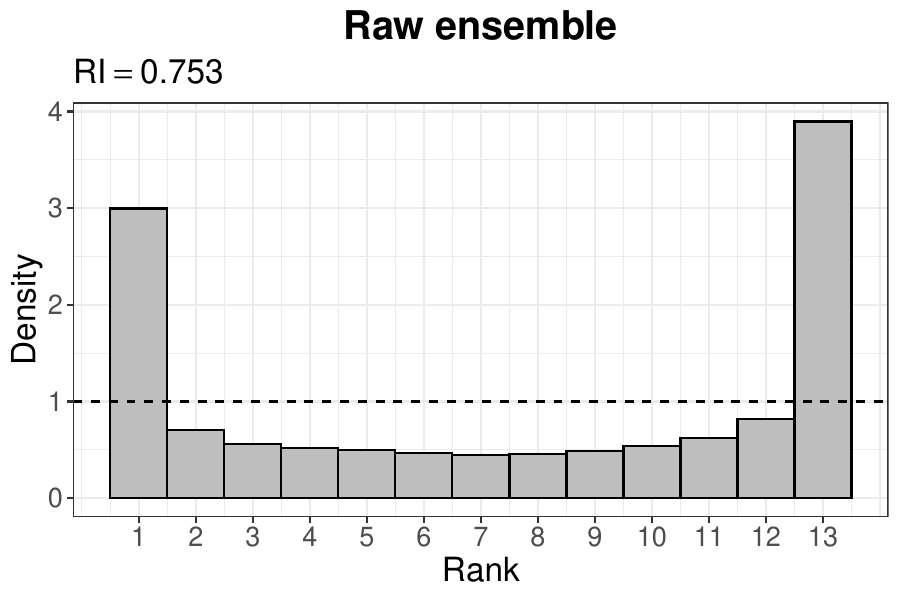}
	\end{center}	
	\caption{Verification rank histogram of the raw ensemble aggregated over all stations and time points in the testing data.}
	\label{fig: vr_hist}
\end{figure}

However, calibration of probabilistic forecasts can also be assessed visually. Given a forecast ensemble, a \textit{verification rank histogram} \parencite{Hamill1997} can be employed, which is a histogram of ranks for the verifying observations with respect to its forecast ensemble, see, e.g. Figure \ref{fig: vr_hist}. For a continuous predictive distribution $F$, a \textit{probability integral transform} (PIT, \cite{Dawid1984}) \textit{histogram} can be used, which is a histogram of PIT values. These values are obtained by evaluating the predictive cumulative distribution function at its verifying observations. Any deviation from uniformity observed in either of these histograms indicates some miscalibration which can also be numerically assessed by the \textit{reliability index} (RI, \cite{Monache2006}). This index quantifies the sum of absolute deviations between the expected relative frequency of a calibrated forecast and the observed relative frequency of the predictive forecast for each bin in a histogram.

\paragraph{Point forecasts.}
In order to verify the accuracy of point forecasts for a predictive distribution $F$, such as the median forecast or the mean forecast, the \textit{mean absolute error} (MAE) and the \textit{root mean squared error} (RMSE) are used as additional scores, respectively \parencite{Gneiting2011b}. 

\paragraph{Forecast comparison.}
Moreover, the relative improvement of a probabilistic forecast $F$ over a reference forecast $F_{\mathrm{ref}}$ in terms of a score $\mathcal{S}$ can be assessed by its \textit{skill score} \parencite{Murphy1973}, which is defined as 
\begin{align}
\mathcal{SS}_F:=1-\frac{\overline{\mathcal{S}}_F}{\overline{\mathcal{S}}_{F_{\mathrm{ref}}}},
\end{align}
where $\overline{\mathcal{S}}_F$ and $\overline{\mathcal{S}}_{F_{\mathrm{ref}}}$ represent the mean scores of $F$ and $F_{\mathrm{ref}}$, respectively.

Furthermore, to assess the statistical significance of the differences in scores among two forecasts, a \textit{Diebold-Mariano test} \parencite{Diebold1995} is conducted in combination with a \textit{Benjamini-Hochberg procedure} \parencite{Benjamini1995} to adjust for multiple station testing. A significance level of $\alpha=0.05$ is chosen for all tests. 

\paragraph{Feature importance.}
Similar to \textcite{Rasp2018} the \textit{feature importance} is evaluated based on the permutation importance \parencite{Breiman2001} employing the CRPS as loss function. At first, only the samples of the covariate of interested are randomly permuted in the testing data. Afterwards, the CRPS differences between the forecasts obtained from the permuted testing data and the forecasts received from the original testing data are calculated.

\section{Results}
\label{sec: Results}

At first, the results of every model for all stations considered in the case study are discussed. Afterwards, the models are compared for a single station in more detail.   

\subsection{General results}

\begin{figure}[!h]
\begin{center}
  	  \includegraphics[scale = 0.5]{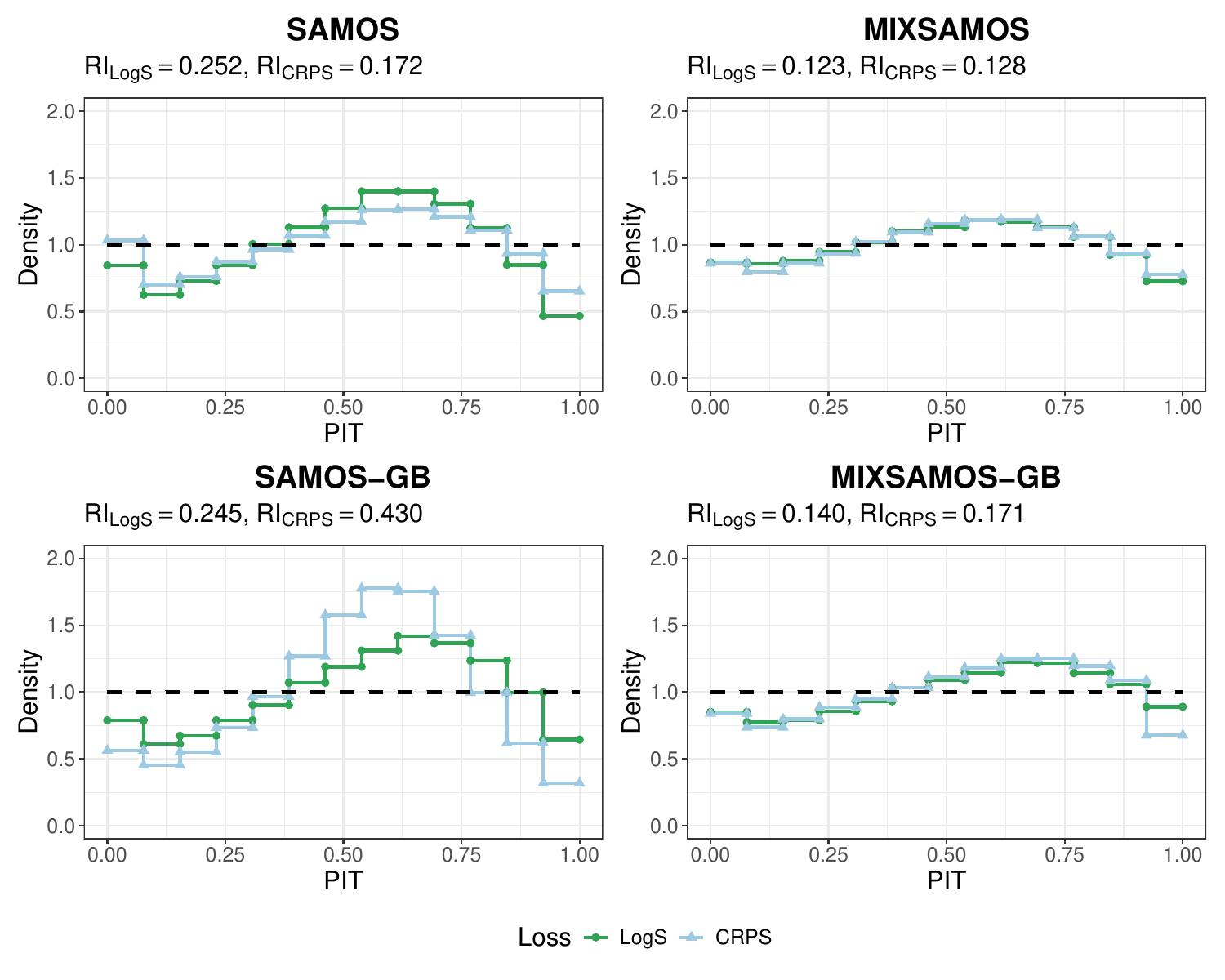}
\end{center}
  	  \caption{PIT histograms of the considered methods including their reliability index (RI), where the PIT values are aggregated over all stations and time points in the testing data.}
      \label{fig: pit_hist}
\end{figure}

Compared to the raw ensemble forecasts, all postprocessing models result into an improved forecast calibration. This can be justified by the flatter PIT histograms in Figure \ref{fig: pit_hist} in contrast to the U-shaped verification rank histogram in Figure \ref{fig: vr_hist}. Although, each postprocessing model successfully corrects the strong underdispersive raw ensemble forecasts, SAMOS(-GB) deliver slightly too underconfident forecasts, resulting in a bump shape in the second half of the PIT histogram, while being a bit too overconfident in the tails of the distributions, see \textcite{Messner2017} for similar results. This pattern becomes especially more distinct for SAMOS-GB if estimated via CRPS instead of LogS as loss function. In contrast, the mixture regression models MIXSAMOS(-GB) reduce this miscalibration caused by a slight distributional misspecifiation and generate the best calibrated forecasts. Consequently, MIXSAMOS(-GB) profit of its distributional flexibility over SAMOS(-GB) which is one reason for its dominating performance. Furthermore, these visual observations are numerically verified by the RI, which is lower and therefore better for the mixture regression models MIXSAMOS(-GB) in comparison to SAMOS(-GB). Furthermore, LogS based optimization results for nearly all models into forecasts with superior calibration in comparison to CRPS based model estimation, which is reflected in a lower RI, too. Therefore, according to the stated paradigm of \textcite{Gneiting2007} in section \ref{sec: Forecast evaluation}, only the methods optimized via LogS are further discussed. For additional results concerning CRPS based model estimation, see Appendix \ref{app: Additional results}.

Table \ref{tab: scores} gives an overview of the predictive performance for all models in terms of the considered evaluation measures in section \ref{sec: Forecast evaluation}. Not surprisingly, all postprocessing models improve upon the raw ensemble. The mixture regression models MIXSAMOS(-GB) outperform their corresponding benchmark model SAMOS(-GB) with respect to CRPS, LogS, MAE and RMSE, respectively. MIXSAMOS-GB shows the strongest enhancements of all methods in terms of the previously mentioned scores. Furthermore, additional covariates help to increase the sharpness of MIXSAMOS-GB upon MIXSAMOS indicated by a lower width, although the forecasts of MIXSAMOS(-GB) are not as sharp as those from their corresponding benchmark models. This loss in sharpness might be related to fact, that the predictive distribution is a combination of two mixture components having independent spread parameters which are driven by different covariates. Consequently, the predictive distribution is at least as dispersed as the least dispersed mixture component \parencite{Gneiting2013}, i.e. the dispersion tends to increase for mixture regression models. This phenomenon is further evidenced in the coverage, where MIXSAMOS(-GB) marginally exceed the nominal coverage of 96.15\%, whereas SAMOS(-GB) align more closely with this target value.

\begin{table}[h!]
\begin{center}
\resizebox{16cm}{!}{
\begin{tabular}{ccccccc}
  \toprule
 & CRPS & LogS & MAE & RMSE & Coverage & Width \\ 
  \hline
Raw ensemble & 1.03 (0.02) & $-$ ($-$) & 1.26 (0.03) & 1.76 (0.04) & 63.09 (0.02) & 2.91 (0.19) \\ \hline
  SAMOS & 0.74 (0.03) & 1.70 (0.03) & 1.00 (0.04) & 1.39 (0.06) & \textbf{96.56} (0.01) & 6.06 (0.20) \\ 
  MIXSAMOS & 0.72 (0.03) & 1.63 (0.03) & 0.99 (0.04) & 1.38 (0.06) & 97.16 (0.01) & 6.51 (0.18) \\ \hline
  SAMOS-GB & 0.71 (0.02) & 1.64 (0.03) & 0.98 (0.03) & 1.33 (0.05) & 96.65 (0.00) & \textbf{5.72} (0.17) \\ 
  MIXSAMOS-GB & \textbf{0.69} (0.02) & \textbf{1.59} (0.03) & \textbf{0.96} (0.03) & \textbf{1.33} (0.05) & 96.83 (0.00) & 5.99 (0.17) \\ 
   \bottomrule
\end{tabular}
}
\end{center}
\caption{Verification scores of all methods aggregated over all stations and time points in the testing data. Bold values represent the best value for each score and the values in brackets denote bootstrap standard errors \parencite{Politis1994}.} 
\label{tab: scores}
\end{table}

\begin{figure}[!h]
\begin{center}
  	  \includegraphics[scale = 0.55]{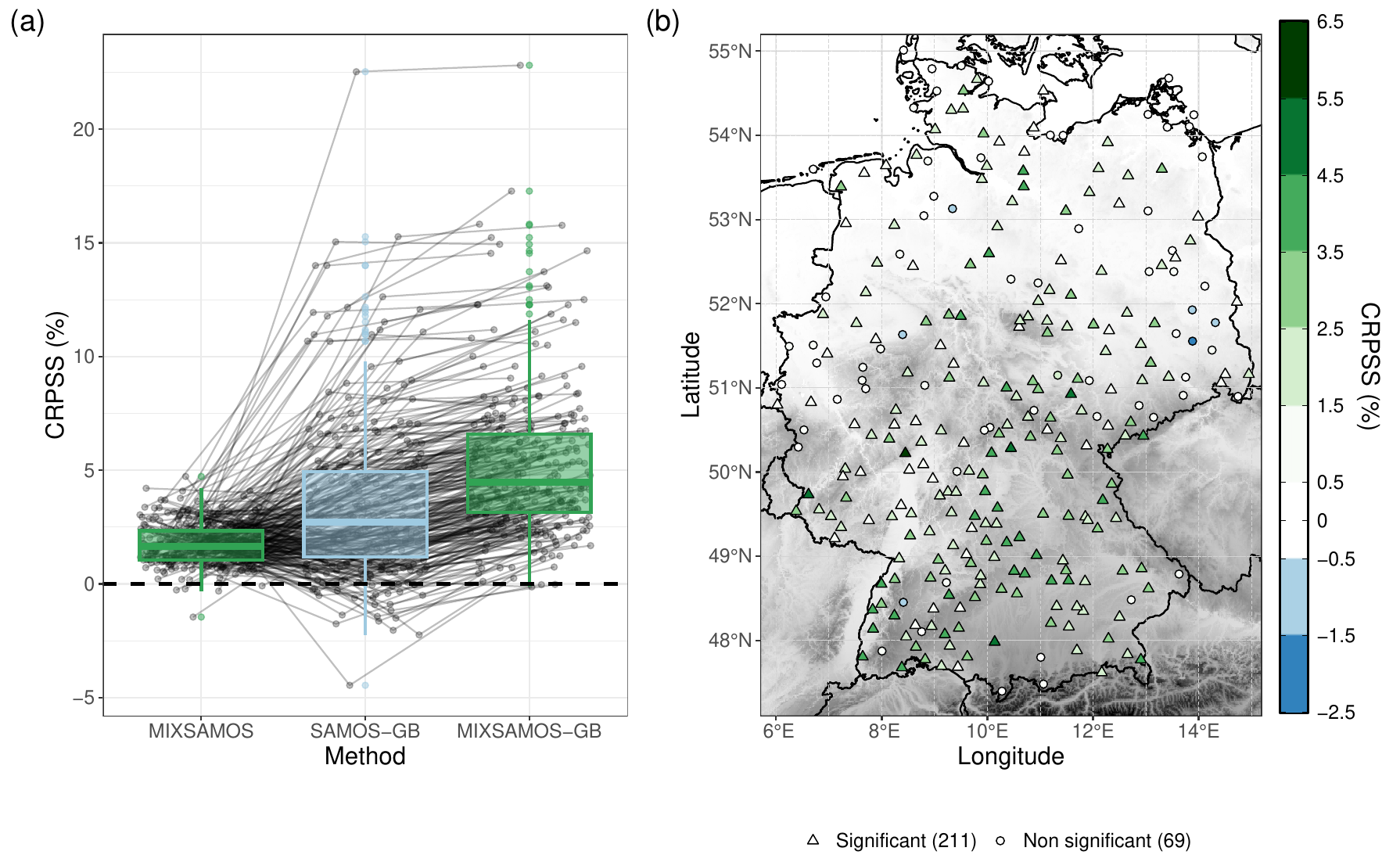}
\end{center}
  	  \caption{(a): boxplots of stationwise CRPS based skill score (CRPSS) improvements (\%) over SAMOS in the testing data. The points show the stationwise CRPSS where the lines interrelate the improvements for the different models. (b): stationwise CRPSS (\%) of MIXSAMOS-GB over SAMOS-GB in the testing data. Significant CRPS differences in favor of MIXSAMOS-GB are indicated by $\triangle$.}
      \label{fig: crpss_bp_map}
\end{figure}

To get better insight into the forecast performance differences among the different models, Figure \ref{fig: crpss_bp_map} (a) shows boxplots of stationwise CRPS based skill score (CRPSS) improvements over SAMOS. Although, MIXSAMOS outperforms SAMOS in median for only around 1.6\%, its CRPS improvements over SAMOS are significant at 86.4\% of all stations. SAMOS-GB enhances SAMOS in median round 2.7\% and shows a higher variance in CRPSS than MIXSAMOS. The strongest improvements are achieved by MIXSAMOS-GB in median around 4.5\%. Additionally, MIXSAMOS-GB shows in comparison to SAMOS-GB nearly no degraded performance over SAMOS, as negative CRPSS are rare. 

Figure \ref{fig: crpss_bp_map} (b) shows the stationwise CRPSS improvements of MIXSAMOS-GB to its direct benchmark model SAMOS-GB. At most of the stations CRPSS gains between $1.5\%-4.5\%$ are possible for MIXSAMOS-GB. Especially in the south of Germany, where 2\,\si{\metre} surface temperature observations exhibit higher uncertainties \parencite{Jobst2024}, stronger CRPSS improvements can be achieved. Moreover, MIXSAMOS-GB significantly outperforms SAMOS-GB in terms of CRPS at around 72.5\% of all stations, indicated by the triangles in Figure \ref{fig: crpss_bp_map} (b). For more details on significant performance differences, please refer to Appendix \ref{app: Additional results}.

\begin{figure}[!h]
\begin{center}
  	  \includegraphics[scale = 0.3]{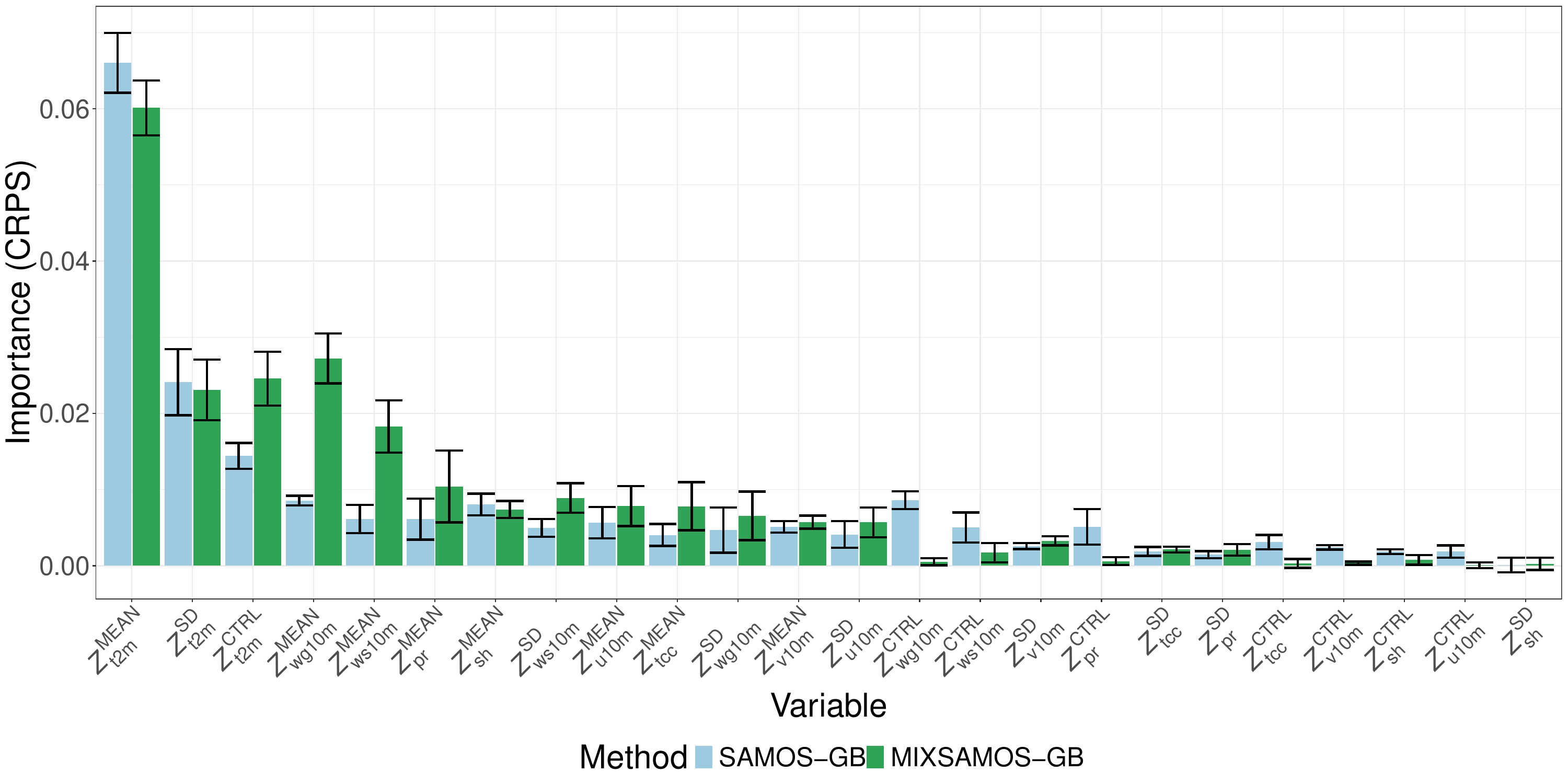}
\end{center}
  	  \caption{CRPS based mean feature importance with bootstrap standard error bars over all stations and time points in the testing data. Importance of $Z_{\mathrm{t2m}}^{\mathrm{MEAN}}$ is divided by 50 for a better representation.}
      \label{fig: vi_crps_logs}
\end{figure}

Assessing feature importance of data driven model is essential as it enables meteorological services to develop parsimonious and tracetable weather forecasting models or to refine existing models accordingly. Figure \ref{fig: vi_crps_logs} shows the CRPS based mean feature importance of all available covariates for SAMOS-GB and MIXSAMOS-GB. First of all, ensemble mean forecasts tend to be more relevant than control forecasts or ensemble standard deviations for both models. Obviously, $Z_{\mathrm{t2m}}^{\mathrm{MEAN}}$ is the most important covariate for both models, although it is slightly more relevant for SAMOS-GB than for MIXSAMOS-GB. Furthermore, $Z_{\mathrm{t2m}}^{\mathrm{SD}}$ is the second most informative feature for both models. In contrast, MIXSAMOS-GB identifies $Z_{\mathrm{t2m}}^{\mathrm{CTRL}}$ to be more important than SAMOS-GB does. This indicates that MIXSAMOS-GB extracts more, possibly useful, information of $Z_{\mathrm{t2m}}^{\mathrm{CTRL}}$ for the predictive distribution than SAMOS-GB, which might be another reason for its superior performance. Similar feature importance differences among both models can be found for e.g. wind related variables $Z_{\mathrm{wg10m}}^{\mathrm{MEAN}}$, $Z_{\mathrm{ws10m}}^{\mathrm{MEAN}}$ which show a comparable importance to $Z_{\mathrm{t2m}}^{\mathrm{SD}}$ for MIXSAMOS-GB. Furthermore, variables which are meteorologically directly related to $Z_{\mathrm{t2m}}^{\mathrm{MEAN}}$, such as $Z_{\mathrm{pr}}^{\mathrm{MEAN}}$ or $Z_{\mathrm{sh}}^{\mathrm{MEAN}}$ appear among the 10 most important features. 

\subsection{Station specific results}

To elucidate the reasons for the superior performance of MIXSAMOS(-GB) over SAMOS(-GB), the results are examined in more detail using an exemplary station called Dillingen/Donau-Fristingen, see Figure \ref{fig: stations}. 

\begin{figure}[!h]
\begin{center}
  	  \includegraphics[scale = 0.4]{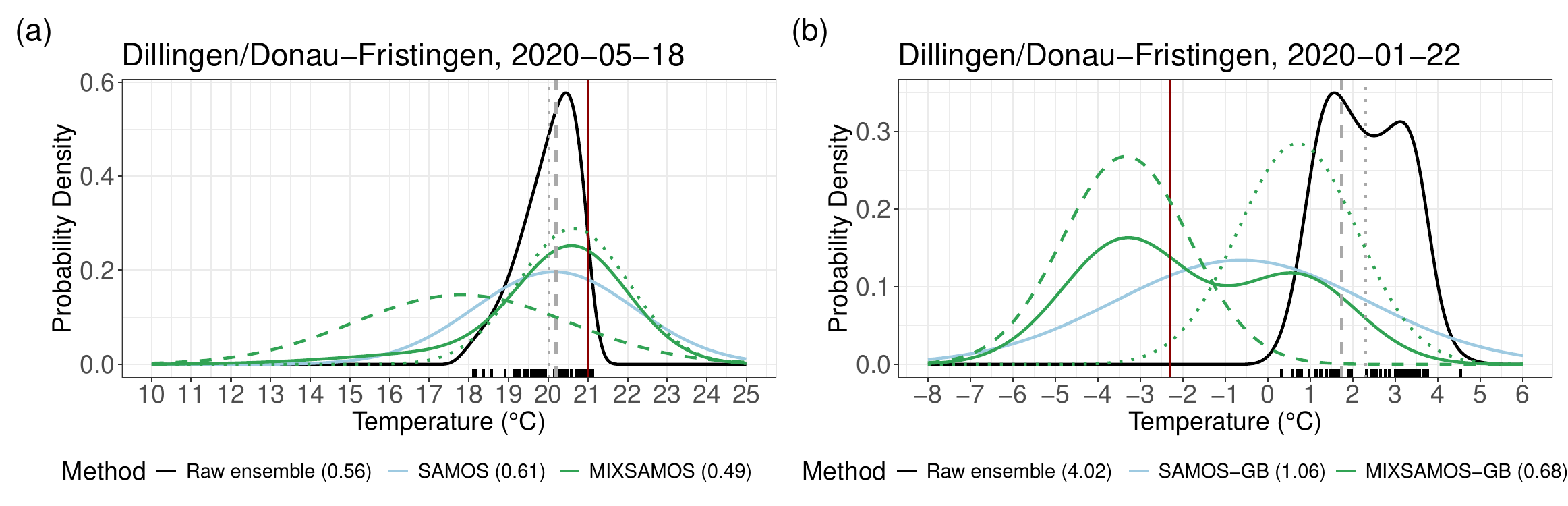}
\end{center}
  	  \caption{2\,\si{\metre} temperature predictive PDF of the raw ensemble (black line), SAMOS(-GB) (lightblue line) and MIXSAMOS(-GB) (green line) with corresponding CRPS in brackets. The red line denotes the observation, the gray dotted and dashed lines represent the deterministic ensemble mean and control forecast, respectively. The green dotted and dashed lines represent the mixture component corresponding to the first $\bm{x}_1$ and second group $\bm{x}_2$ of MIXSAMOS(-GB), respectively.}
      \label{fig: density_plots}
\end{figure}

Figure \ref{fig: density_plots} shows the predictive PDFs of the raw ensemble, SAMOS(-GB) and MIXSAMOS(-GB) at Dillingen/Donau-Fristingen. MIXSAMOS(-GB) clearly outperform the raw ensemble and SAMOS(-GB) with respect to CRPS on both days. In panel (a) of Figure \ref{fig: density_plots}, the predictive PDF of MIXSAMOS is slightly left-skewed, reflecting the skewness contained in the raw ensemble, in order to catch the observation. The skewness of the predictive PDF of MIXSAMOS is mainly caused by the mixture component of the second group $\bm{x}_2=(x_{\mathrm{t2m}}^{\mathrm{CTRL}})$, which underforecasts the observation in contrast to the mixture component of the first group $\bm{x}_1=(x_{\mathrm{t2m}}^{\mathrm{MEAN}}, x_{\mathrm{t2m}}^{\mathrm{SD}})$. Additionally, the mixture component corresponding the second group shows a higher spread in contrast to one of the first group. This is primarily due to the fact that the scale parameter of the second mixture component solely relies on the intercept and not on any other covariates. In contrast, the scale parameter of the first mixture component depends on the ensemble standard deviation containing relevant uncertainty information. Thus, the spread of the mixture component for the first group is more consistent with the ensemble spread, resulting in a sharper predictive PDF, than the one related to the second group. In panel (b) of Figure \ref{fig: density_plots}, the predictive PDF of MIXSAMOS-GB adopts the bimodality of the raw ensemble, which enables MIXSAMOS-GB to assign a higher density to the region of the actual observation than SAMOS-GB. In this example, the control forecast is slightly closer to the observation than the ensemble mean forecast. Compared to the scenario in panel (a), MIXSAMOS-GB conserves this forecast order, as the mixture component of the second group is closer to the observation than the one of the first group. In contrast to panel (a), the mixture component of the second group has now a comparable spread to the one of the first group, influenced by the control forecasts of additional weather variables. Eventually, the mixture component of MIXSAMOS(-GB) that aligns more closely with the observation is allocated the larger of the two mixture weights in both panels. This signifies that MIXSAMOS(-GB) possess the capability to identify the mixture component that provides a superior fit.

\begin{figure}[!h]
\begin{center}
  	  \includegraphics[scale = 0.42]{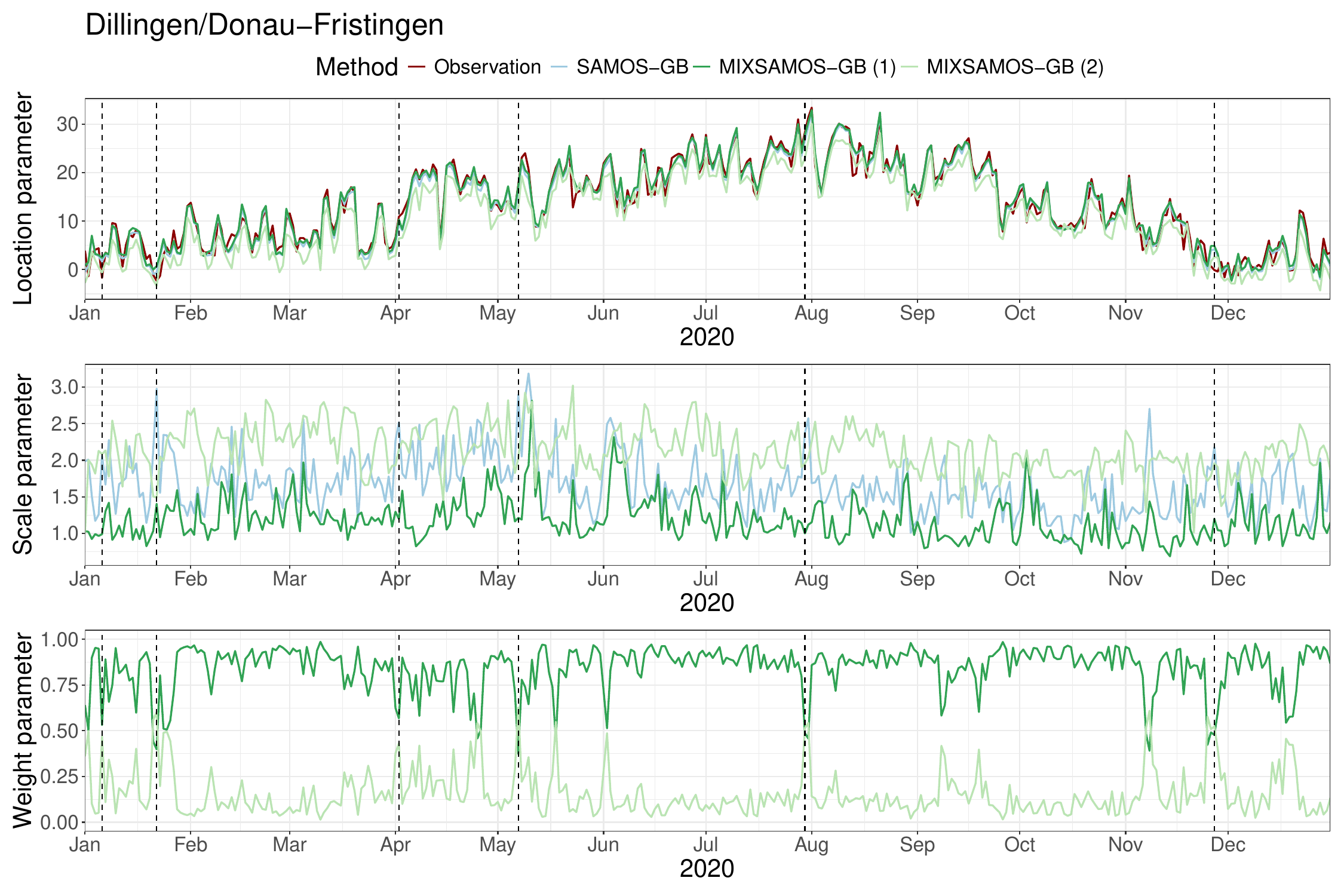}
\end{center}
  	  \caption{Estimated parameters of SAMOS-GB and MIXSAMOS-GB for the testing data.}
      \label{fig: parameter_ts}
\end{figure}

The estimated parameters of SAMOS-GB and of MIXSAMOS-GB for the testing data are visualized in Figure \ref{fig: parameter_ts}. The location parameter of SAMOS-GB and the location parameter of the first group of MIXSAMOS-GB align for most forecast cases well with the observation. However, the location parameter corresponding to the second group of MIXSAMOS-GB systematically underforecasts the observation. Consequently, a higher uncertainty, indicated by a larger scale parameter is assigned to the second group in contrast to the one of the first group for MIXSAMOS-GB. In the case of abrupt temperature drops the location parameter of the first mixture component may overforecast the observation, while the location parameter of the second mixture component may be closer to the observation due to tendency for underforecasting, see, e.g. dates 2020-01-06, 2020-01-22, 2020-11-27. Consequently, MIXSAMOS-GB drastically increases the generally lower weight for the second mixture component, as learned from the longer static training data, in order to better match the predictive distribution with the observation. However, also in the case of a sudden temperature rise, the weight corresponding to the second mixture component is increased in order to account for the potentially necessary uncertainty, see, e.g. dates 2020-04-02, 2020-05-07, 2020-07-30. 

\begin{figure}[!h]
\begin{center}
  	  \includegraphics[scale = 0.25]{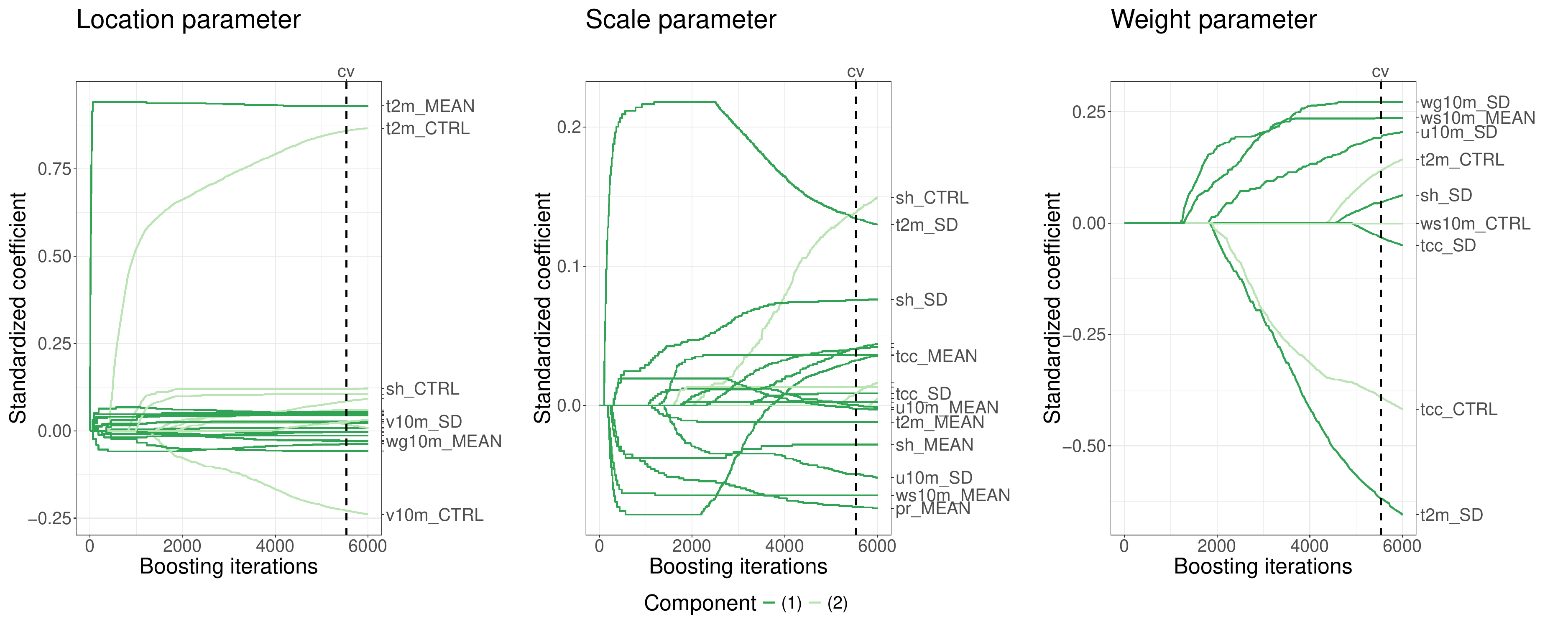}
\end{center}
  	  \caption{Coefficient paths for the location, scale and weight parameters of MIXSAMOS-GB. The vertical dashed lines shows $m_{\mathrm{opt}}=5536$ determined via CV.}
      \label{fig: coef_path}
\end{figure}

While the joint variable importance for all parameters of all stations is presented in Figure \ref{fig: vi_crps_logs}, the coefficient paths of the selected (standardized) covariates are illustrated for each distribution parameter of MIXSAMOS-GB in Figure \ref{fig: coef_path}. After initializing all coefficients with zero, $Z_{\mathrm{t2m}}^{\mathrm{MEAN}}$ and $Z_{\mathrm{t2m}}^{\mathrm{CTRL}}$ are the first  covariates for both location parameters receiving a non-zero coefficient. During the boosting process additional covariates are included for the location parameter of the first mixture component, e.g. $Z_{\mathrm{sh}}^{\mathrm{MEAN}}$, $Z_{\mathrm{ws10m}}^{\mathrm{MEAN}}$ and of the second mixture component, e.g. $Z_{\mathrm{sh}}^{\mathrm{CTRL}}$, $Z_{\mathrm{v10m}}^{\mathrm{CTRL}}$. After the optimal number of boosting iterations is found at $m_{\mathrm{opt}}=5536$, $Z_{\mathrm{t2m}}^{\mathrm{MEAN}}$ and $Z_{\mathrm{t2m}}^{\mathrm{CTRL}}$ have the largest coefficient with respect to its corresponding location parameter, indicating that these covariates are the most informative as expected. The variable $Z_{\mathrm{sh}}^{\mathrm{CTRL}}$ is the most important covariate for the scale parameter of the second mixture component, while only a few covariates are selected for the scale parameter of the second mixture component. Moreover, the variable  $Z_{\mathrm{t2m}}^{\mathrm{SD}}$ is the most relevant covariate for the scale parameter of the first mixture component, but its coefficient slightly decreases after approximately 2500 iterations. One reason for this behavior might be, that the variance is better explained by the meanwhile additional selected covariates, such as e.g. $Z_{\mathrm{sh}}^{\mathrm{SD}}$, $Z_{\mathrm{pr}}^{\mathrm{MEAN}}$. Another reason might be the selection of $Z_{\mathrm{t2m}}^{\mathrm{SD}}$ as covariate for the mixture weight of the first mixture component at around 2000 iterations. While the coefficient of $Z_{\mathrm{tcc}}^{\mathrm{MEAN}}$ for the scale parameter of the first mixture component increases again after around 2500 iterations and becomes positive, the coefficient of $Z_{\mathrm{tcc}}^{\mathrm{CTRL}}$ clearly decreases for the second mixture weight and becomes its most important one. After $m_{\mathrm{opt}}$ is reached, $Z_{\mathrm{t2m}}^{\mathrm{SD}}$ is the most relevant covariate for the first mixture weight followed by others, such as $Z_{\mathrm{wg10m}}^{\mathrm{SD}}$, $Z_{\mathrm{ws10m}}^{\mathrm{MEAN}}$. Consequently, the selection of covariates $Z_{\mathrm{t2m}}^{\mathrm{SD}}$, $Z_{\mathrm{tcc}}^{\mathrm{CTRL}}$ for the mixture weights influences the coefficient path for covariates $Z_{\mathrm{t2m}}^{\mathrm{SD}}$, $Z_{\mathrm{tcc}}^{\mathrm{MEAN}}$ in the location parameter of the first mixture component. Eventually, Figures \ref{fig: parameter_ts} and \ref{fig: coef_path} indicate that a non-constant mixture weight modeling provides valuable information regarding the predictive performance and relevance of the different exchangeable groups and covariates.

\section{Conclusion and outlook}
\label{sec: Conclusion and outlook}

In this work, the non-cyclic gradient-boosting algorithm for mixture regression models is presented. Besides an intrinsic variable selection, this algorithm estimates regression coefficients corresponding to the (selected) covariates. Furthermore, the algorithm shrinks regression coefficients in order to avoid overfitting, resulting in superior fits. To enhance implementation and adaptability to a broader range of applications, the employed algorithms are provided in the \texttt{R}-package \texttt{mixnhreg} \parencite{Jobst2024c}.

For the postprocessing of ensemble weather forecasts, more general mixture regression models are proposed. MIXMOS serves as baseline model where the ensemble forecasts of an exchangeable group are only linked to the distribution parameters and the mixture weight of a single mixture component. To address seasonal effects, MIXSAMOS utilizes standardized anomalies rather than relying on the original data, as is the case with MIXMOS. Additionally, MIXSAMOS can be estimated via the previously introduced non-cyclic gradient-boosting algorithm, which provides a first automatic variable selection for mixture regression models in this application field. Moreover, the gradients with respect to the predictors considering the CRPS loss are derived, see Appendix \ref{app: Gradients for the mixture normal distribution}. 

The proposed mixture regression models are applied in a case study for postprocessing 2\,\si{\metre} surface temperature at 280 observation stations in Germany. The novel mixture regression models MIXSAMOS(-GB) provide forecasts that are more accurately calibrated, than the ones of the benchmark models SAMOS(-GB). Furthermore, MIXSAMOS(-GB), demonstrate superior performance compared to their corresponding benchmark models SAMOS(-GB) across nearly all scores. Notably, MIXSAMOS-GB significantly outperforms all alternative models with respect to CRPS. One reason for the overall superior performance of the mixture regression models is, that they are able to integrate distributional properties of the raw ensemble such as, e.g. skewness or multimodality into the predictive distribution. Furthermore, non-constant mixture weights result into a more adaptive  predictive distribution being able to account for e.g. sudden weather changes. Eventually, the gradient-boosting algorithm of MIXSAMOS-GB selects the most relevant covariates for all mixture weights and distribution parameters resulting in explainable weather forecasting models with enhanced forecast skill. Although, only two mixture components from a single distribution family are used for MIXSAMOS(-GB) in the case study, it is readily feasible to extend these models to more than two mixture components, potentially from various distribution families. 

One limitation of the proposed mixture regression models is, that they do not take account of spatial relationships. Additionally, these models are not well suited to generate forecasts at unobserved locations, as they are estimated locally. Therefore, besides removing temporal effects, also spatial effects can be addressed in the climatologies similar to \textcite{Dabernig2017} leading to spatio-temporal adaptive mixture regression models. Furthermore, the novel mixture regression models can be straightforwardly adapted to the simultaneous ensemble postprocessing of various lead times \parencite{Dabernig2017a}. Moreover, only small modifications of the considered distribution families are necessary in order to adjust the mixture regression models to the ensemble postprocessing of other weather variables such as wind speed or precipitation. Finally, it would be worthwhile to examine these models for postprocessing multi-model ensemble forecasts to discern the strengths and weaknesses of the forecasts generated by different forecasting systems. All these ideas demonstrate, that there is still lots of research potential left to further improve mixture regression models in the field of ensemble postprocessing.

\section*{Acknowledgements}
The author is grateful to the European Centre for Medium-Range Weather Forecasts (ECMWF) and the German Weather Service (DWD) for providing forecasts and observation data, respectively. Furthermore, the author acknowledges support of the research by Deutsche Forschungsgemeinschaft (DFG) Grant Number 395388010.

\clearpage
\pagebreak
\newpage
\appendix
\begin{Large}
\textbf{Appendix}	
\end{Large}

\section{Additional results}
\label{app: Additional results}

\begin{table}[h!]
\begin{center}
\resizebox{16cm}{!}{
\begin{tabular}{ccccccc}
  \toprule
 & CRPS & LogS & MAE & RMSE & Coverage & Width \\ 
  \hline
Raw ensemble & 1.03 (0.02) & $-$ ($-$) & 1.26 (0.03) & 1.76 (0.04) & 63.09 (0.02) & 2.91 (0.19) \\ \hline
  SAMOS & 0.73 (0.03) & 1.70 (0.05) & 0.99 (0.04) & 1.38 (0.06) & 95.19 (0.01) & \textbf{5.45} (0.18) \\ 
 MIXSAMOS & 0.72 (0.03) & 1.65 (0.03) & 0.99 (0.04) & 1.38 (0.06) & \textbf{96.31} (0.01) & 6.07 (0.18) \\ \hline
SAMOS-GB & 0.76 (0.02) & 1.79 (0.03) & 0.98 (0.03) & 1.34 (0.05) & 97.61 (0.01) & 7.42 (0.24) \\ 
MIXSAMOS-GB & \textbf{0.70} (0.03) & \textbf{1.62} (0.03) & \textbf{0.96} (0.03) & \textbf{1.32} (0.06) & 96.87 (0.01) & 6.11 (0.19) \\ 
   \bottomrule
\end{tabular}
}
\end{center}
\caption{Verification scores of all methods optimized via CRPS aggregated over all stations and time points in the testing data. Bold values represent the best value for each score and the values in brackets denote bootstrap standard errors \parencite{Politis1994}.} 
\end{table}

\begin{figure}[!h]
\begin{center}
  	  \includegraphics[scale = 0.4]{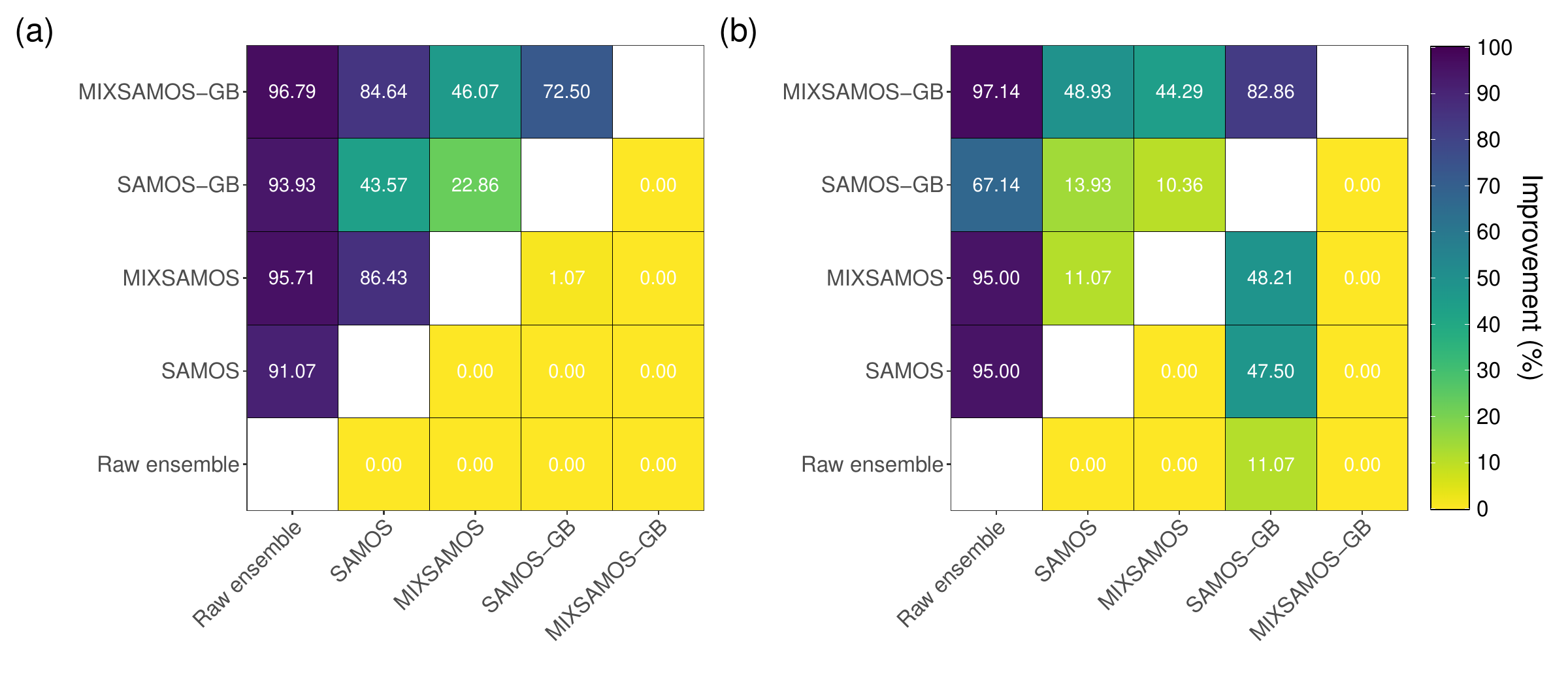}
\end{center}
  	  \caption{Percentage of stations, where pair-wise Diebold-Mariano (DM) tests indicate statistically significant CRPS improvements of the method in the row over the method in the column after applying the Benjamini-Hochberg procedure.\\ (a): all methods are optimized via LogS. (b): all methods are optimized via CRPS.}
      \label{fig: sig_crps}
\end{figure}

\begin{figure}[!h]
\begin{center}
  	  \includegraphics[scale = 0.3]{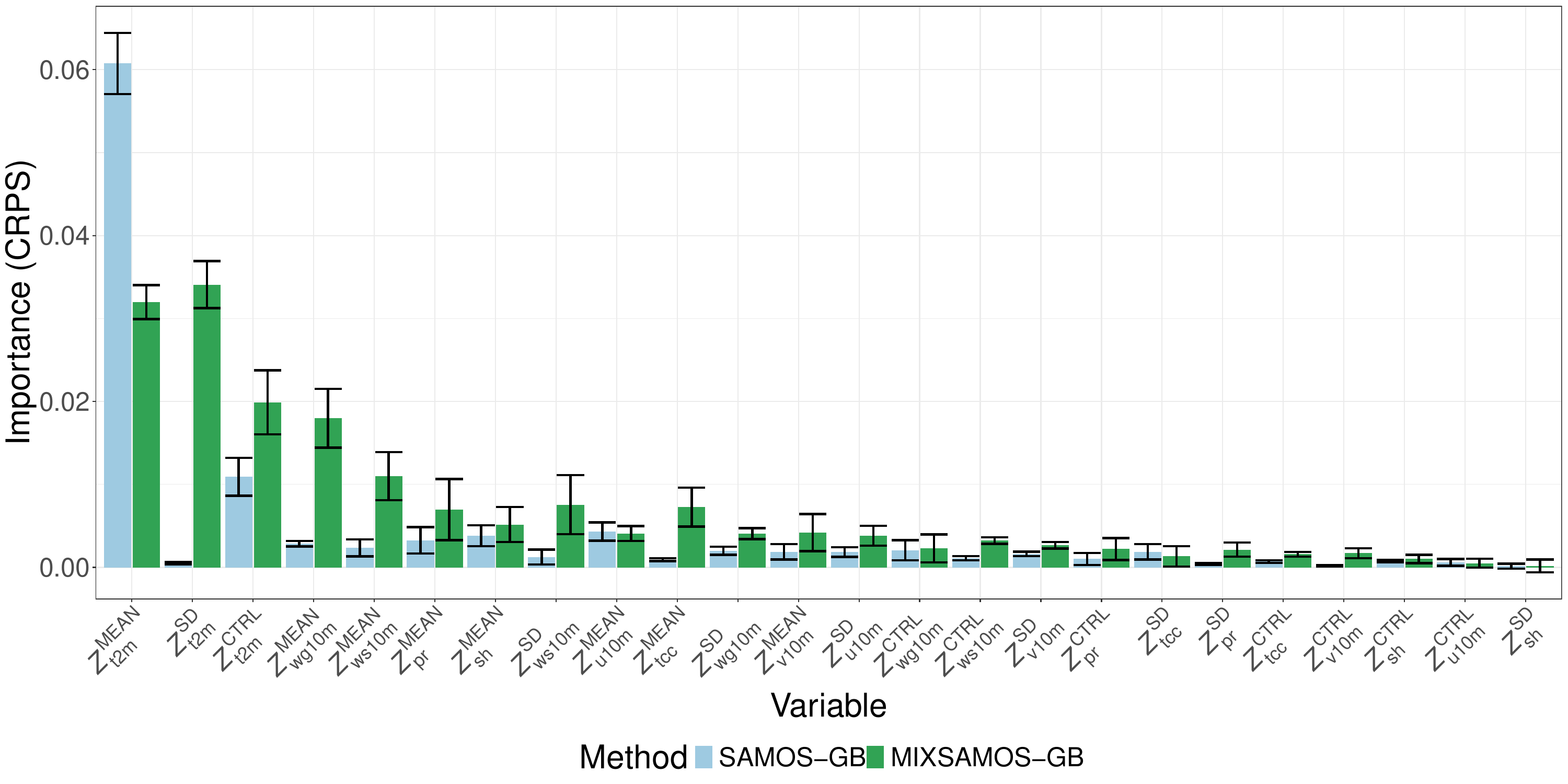}
\end{center}
  	  \caption{CRPS based mean feature importance with bootstrap standard error bars over all stations and time points in the testing data. Importance of $Z_{\mathrm{t2m}}^{\mathrm{MEAN}}$ is divided by 50 and $Z_{\mathrm{t2m}}^{\mathrm{CTRL}}$ is divided by 10 for a better representation. All methods are optimized via CRPS.}
      \label{fig: vi_crps_crps}
\end{figure}
\pagebreak

\section{Hyperparameters}
\label{app: Hyperparameters}

\begin{table}[h!]
\begin{center}
\begin{tabular}{l l l} 
\toprule
\textbf{Method} & Hyperparameter & Value\\ \hline 
\textbf{SAMOS} & Maximum number of iterations & 5000 \\
& Relative threshold & $1\mathrm{e}^{-8}$\\ \hline 
\textbf{SAMOS-GB} & Maximum number of iterations & 2000 \\
& Step size & 0.05 \\
& Stopping criterion & 10-fold CV\\ \hline 
\textbf{MIXSAMOS-GB} & Maximum number of iterations & 6000 \\
& Step size & 0.05 \\
& Stopping criterion & 10-fold CV\\  
\bottomrule
\end{tabular}
\end{center}
\caption{Overview of the hyperparameter specifications.}
\label{tab: overview_config}
\end{table}

\pagebreak
\section{Gradients for the mixture normal distribution}
\label{app: Gradients for the mixture normal distribution}

Subsequently, the mixture regression model 
\begin{align}
f(y\,\vert\,\bm{x}):=\sum\limits_{k=1}^{K}\omega_k(\bm{x})f_k(y\,\vert\, \mu_k(\bm{x}), \sigma_k(\bm{x})^2), \label{eq: MRM_app}
\end{align}
where each mixture component follows a normal distribution, i.e. $f_k\sim \mathcal{N}(\mu_k(\bm{x}), \sigma_k(\bm{x})^2)$, is considered. To ensure the requirements for the mixture weights, a softmax function is employed to link the first $K$ linear predictors 
\begin{align*} \omega_k(\bm{x}):=\frac{\exp(\eta_{\omega_k}(\bm{x}))}{\sum\limits_{j=1}^{K}\exp(\eta_{\omega_j}(\bm{x}))},\quad k=1,\ldots,K,
\end{align*}
to the mixture weights. Additionally, the inverse link functions for the location parameters are given by $g_{\mu_k}^{-1}:=\mathrm{id}$ and for the scale parameters by $g_{\sigma_k}^{-1}:=\log$ for all $K$ mixture components. To simplify notation, the covariates $\bm{x}$ linked to the distribution parameters are omitted for the following statements.\\

\textbf{Statement 1} \parencite{Bishop1994}\\
Let a mixture regression model be given as in Equation \eqref{eq: MRM_app}. Then, the gradients of the loss function $\ell=\mathrm{LogS}$ with respect to the linear predictors are given by
\begin{align*}
    \frac{\partial \ell}{\partial \eta_{\mu_k}}&=\pi_k\cdot\left(\frac{\mu_k-y}{\sigma_k^2}\right),\\
    \frac{\partial \ell}{\partial \eta_{\sigma_k}}&=\pi_k\cdot \left(1-\left(\frac{\mu_k-y}{\sigma_k}\right)^2\right),\\
    \frac{\partial \ell}{\partial \eta_{\omega_k}}&=\omega_k-\pi_k,
\end{align*}
with posterior probabilities
\begin{align*}
\pi_k:=\frac{\omega_kf_k(y\,\vert\,\mu_k, \sigma_k^2)}{\sum\limits_{i=1}^{K} \omega_if_i(y\,\vert\,\mu_i, \sigma_i^2)},\quad k=1,\ldots,K.
\end{align*}

\textbf{Proof:}\\
This proof is based on the more general version by \textcite{Bishop1994}. Due to the inverse link function $g_{\mu_k}^{-1}:=\mathrm{id}$, it follows that 
\begin{align*}
\frac{\partial \eta_{\mu_k}}{\partial \mu_k}=\frac{\partial \mu_k}{\partial \mu_k}=1\quad \Leftrightarrow\quad \frac{\partial \mu_k}{\partial \eta_{\mu_k}}=1. 
\end{align*}
By employing elementary rules for derivatives, one gets
\begin{align*}
    \frac{\partial \ell}{\partial \eta_{\mu_k}}&=\frac{\partial \mu_k}{\partial \eta_{\mu_k}}\frac{\partial \ell}{\partial \mu_k}=\frac{\partial \ell}{\partial \mu_k}=-\frac{\omega_kf_k(y\,\vert\,\mu_k, \sigma_k^2)}{\sum\limits_{i=1}^{K} \omega_if_i(y\,\vert\,\mu_i, \sigma_i^2)}\cdot \left(\frac{y-\mu_k}{\sigma_k^2}\right)=\pi_k\cdot\left(\frac{\mu_k-y}{\sigma_k^2}\right).
\end{align*}
As a result of the inverse link function $g_{\sigma_k}^{-1}:=\log$, it can be concluded that
\begin{align*}
\frac{\partial \eta_{\sigma_k}}{\partial \sigma_k}=\frac{\partial \log(\sigma_k)}{\partial \sigma_k}=\frac{1}{\sigma_k}\quad \Leftrightarrow\quad \frac{\partial \sigma_k}{\partial \eta_{\sigma_k}}=\sigma_k.
\end{align*}
Again, using elementary rules for derivatives yields to
\begin{align*}
    \frac{\partial \ell}{\partial \eta_{\sigma_k}}&=\frac{\partial \sigma_k}{\partial \eta_{\sigma_k}}\frac{\partial \ell}{\partial \sigma_k}=\sigma_k\left[-\frac{\omega_kf_k(y\,\vert\,\mu_k, \sigma_k^2)}{\sum\limits_{i=1}^{K} \omega_if_i(y\,\vert\,\mu_i, \sigma_i^2)}\cdot \left(\frac{(y-\mu_k)^2}{\sigma_k^3}-\frac{1}{\sigma_k}\right)\right]=\pi_k\cdot \left(1-\left(\frac{\mu_k-y}{\sigma_k}\right)^2\right).
\end{align*}
In order to derive the gradients with respect to the linear predictors of the mixture weights, we first observe for $j\neq k$
\begin{align*}
\frac{\partial \omega_j}{\partial \eta_{\omega_k}}=-\frac{\exp(\eta_{\omega_j})\exp(\eta_{\omega_k})}{\left(\sum\limits_{i=1}^{K}\exp(\eta_{\omega_i})\right)^2}=-\omega_j\omega_k,
\end{align*}
and for $j=k$
\begin{align*}
\frac{\partial \omega_j}{\partial \eta_{\omega_k}}=\frac{\exp(\eta_{\omega_k})\sum\limits_{i=1}^{K}\exp(\eta_{\omega_i}) - \exp(\eta_{\omega_k})\exp(\eta_{\omega_k})}{\left(\sum\limits_{i=1}^{K}\exp(\eta_{\omega_i})\right)^2}=\omega_k-\omega_k^2. 
\end{align*}
To sum up, one receives 
\begin{align}
\frac{\partial \omega_j}{\partial \eta_{\omega_k}}=\delta_{jk}\omega_j-\omega_j\omega_k, \label{eq: grad_weight1}
\end{align}
where $\delta_{jk}$ denotes the Kronecker delta. Consequently, the gradients with respect to the linear predictors of the mixture weights are given via
\begin{align*}
    \frac{\partial \ell}{\partial \eta_{\omega_k}}&=-\frac{\displaystyle \frac{\partial}{\partial \eta_{\omega_k}}\sum_{i=1}^{K} \omega_if_i(y\,\vert\,\mu_i, \sigma_i^2)}{\sum\limits_{i=1}^{K} \omega_if_i(y\,\vert\,\mu_i, \sigma_i^2)}=-\frac{\displaystyle \sum_{i=1}^{K}(\delta_{jk}\omega_j-\omega_j\omega_k)f_i(y\,\vert\,\mu_i, \sigma_i^2)}{\sum\limits_{i=1}^{K} \omega_if_i(y\,\vert\,\mu_i, \sigma_i^2)}\\
    &=-\frac{-\omega_k\sum\limits_{i=1}^{K} \omega_if_i(y\,\vert\,\mu_i, \sigma_i^2) + \omega_kf_k(y\,\vert\,\mu_k, \sigma_k^2)}{\sum\limits_{i=1}^{K} \omega_if_i(y\,\vert\,\mu_i, \sigma_i^2)}=\omega_k-\pi_k.
\end{align*}
\begin{flushright}
$\Box$
\end{flushright}
\pagebreak

\textbf{Statement 2}\\
Let a mixture regression model be given as in Equation \eqref{eq: MRM_app}. Then, the gradients of the loss function $\ell=\mathrm{CRPS}$ with respect to the linear predictors are given by
\begin{align*}
    \frac{\partial \ell}{\partial \eta_{\mu_k}}&=\omega_k\left[1-2\Phi\left(\frac{y-\mu_k}{\sigma_k}\right)+
    \sum\limits_{i=1}^{K}\omega_i\left(1-2\Phi\left(\frac{\mu_k-\mu_i}{\sqrt{\sigma_k^2+\sigma_i^2}}\right)\right)\right],\\
    \frac{\partial \ell}{\partial \eta_{\sigma_k}}&=2\omega_k\sigma_k\left[\varphi\left(\frac{y-\mu_k}{\sigma_k}\right)-\sum\limits_{i=1}^{K}\omega_i\frac{\sigma_k}{\sqrt{\sigma_k^2+\sigma_i^2}}\varphi\left(\frac{\mu_k-\mu_i}{\sqrt{\sigma_k^2+
    \sigma_i^2}}\right)\right],\\
    \frac{\partial \ell}{\partial \eta_{\omega_k}}&=-2\omega_k\mathrm{CRPS}(F,y)+\sum\limits_{i=1}^{K}(\delta_{ik}\omega_i-\omega_i\omega_k)A(y-\mu_i, \sigma_i^2)-\omega_k\sum\limits_{i=1}^{K}\omega_iA(\mu_i-\mu_k, \sigma_i^2+\sigma_k^2).
\end{align*}

\textbf{Proof:}\\
Due to the symmetry of $\varphi$ and $\Phi(z)=1-\Phi(-z)$ for $z\in \R$ ($\ast$), Equation \eqref{eq: CRPS2} can be written as 
\begin{footnotesize}
\begin{align}
    \mathrm{CRPS}(F,y)&=\sum\limits_{i=1}^{K}\omega_iA(y-\mu_i,\sigma_i^2)-\frac{1}{2}\sum\limits_{i=1}^{K}\omega_i^2A(0, 2\sigma_i^2)-\sum\limits_{1\leq i< j \leq K}\omega_i\omega_jA(\mu_i-\mu_j,\sigma_i^2+\sigma_j^2)\label{eq: CRPS3}\\
    &=\sum\limits_{i=1}^{K}\omega_i\left[(y-\mu_i)\left(2\Phi\left(\frac{y-\mu_i}{\sigma_i}\right)-1\right)+2\sigma_i\varphi\left(\frac{y-\mu_i}{\sigma_i}\right)\right]-\sqrt{2}\varphi(0)\sum\limits_{i=1}^{K}\omega_i^2\sigma_i\notag\\
    &-\sum\limits_{1\leq i < j \leq K}\omega_i\omega_j\left[(\mu_i-\mu_j)\left(2\Phi\left(\frac{\mu_i-\mu_j}{\sqrt{\sigma_i^2+\sigma_j^2}}\right)-1\right)+2\sqrt{\sigma_i^2+
    \sigma_j^2}\varphi\left(\frac{\mu_i-\mu_j}{\sqrt{\sigma_i^2+\sigma_j^2}}\right)\right].\notag
\end{align}
\end{footnotesize}By deriving $\frac{\partial \mu_k}{\partial \eta_{\mu_k}}=1$ as in Statement 1 and using elementary rules for derivatives as well as property ($\ast$), one gets
\begin{footnotesize}
\begin{align*}
    &\frac{\partial \ell}{\partial \eta_{\mu_k}}=\frac{\partial \mu_k}{\partial \eta_{\mu_k}}\frac{\partial \ell}{\partial \mu_k}=\frac{\partial \ell}{\partial \mu_k}=\omega_k\left[1-2\Phi\left(\frac{y-\mu_k}{\sigma_k}\right)-2\frac{y-\mu_k}{\sigma_k}\varphi\left(\frac{y-\mu_k}{\sigma_k}\right)+2\frac{y-\mu_k}{\sigma_k}\varphi\left(\frac{y-\mu_k}{\sigma_k}\right)\right]\\
    &-\sum\limits_{l=1}^{k-1}\omega_l\omega_k\left[1-2\Phi\left(\frac{\mu_l-\mu_k}{\sqrt{\sigma_l^2+\sigma_k^2}}\right)-2\frac{\mu_l-\mu_k}{\sqrt{\sigma_l^2+\sigma_k^2}}\varphi\left(\frac{\mu_l-\mu_k}{\sqrt{\sigma_l^2+\sigma_k^2}}\right)+2\frac{\mu_l-\mu_k}{\sqrt{\sigma_l^2+\sigma_k^2}}\varphi\left(\frac{\mu_l-\mu_k}{\sqrt{\sigma_l^2+\sigma_k^2}}\right)\right]\\
    &-\sum\limits_{l=k+1}^{K}\omega_k\omega_l\left[2\Phi\left(\frac{\mu_k-\mu_l}{\sqrt{\sigma_k^2+\sigma_l^2}}\right)-1+2\frac{\mu_k-\mu_l}{\sqrt{\sigma_k^2+\sigma_l^2}}\varphi\left(\frac{\mu_k-\mu_l}{\sqrt{\sigma_k^2+\sigma_l^2}}\right)-2\frac{\mu_k-\mu_l}{\sqrt{\sigma_k^2+\sigma_l^2}}\varphi\left(\frac{\mu_k-\mu_l}{\sqrt{\sigma_k^2+\sigma_l^2}}\right)\right]\\
    &=\omega_k\left(1-2\Phi\left(\frac{y-\mu_k}{\sigma_k}\right)\right)-\sum\limits_{l=1}^{k-1}\omega_l\omega_k\left(1-2\Phi\left(\frac{\mu_l-\mu_k}{\sqrt{\sigma_l^2+\sigma_k^2}}\right)\right)+\sum\limits_{l=k+1}^{K}\omega_k\omega_l\left(1-2\Phi\left(\frac{\mu_k-\mu_l}{\sqrt{\sigma_k^2+\sigma_l^2}}\right)\right)\\
    &=\omega_k\left(1-2\Phi\left(\frac{y-\mu_k}{\sigma_k}\right)\right)+\sum\limits_{l=1}^{k-1}\omega_k\omega_l\left(1-2\Phi\left(\frac{\mu_k-\mu_l}{\sqrt{\sigma_k^2+\sigma_l^2}}\right)\right)+\sum\limits_{l=k+1}^{K}\omega_k\omega_l\left(1-2\Phi\left(\frac{\mu_k-\mu_l}{\sqrt{\sigma_k^2+\sigma_l^2}}\right)\right)\\
    &=\omega_k\left[1-2\Phi\left(\frac{y-\mu_k}{\sigma_k}\right)+
    \sum\limits_{i=1}^{K}\omega_i\left(1-2\Phi\left(\frac{\mu_k-\mu_i}{\sqrt{\sigma_k^2+\sigma_i^2}}\right)\right)\right].
\end{align*}
\end{footnotesize}Calculating $\frac{\partial \sigma_k}{\partial \eta_{\sigma_k}}=\sigma_k$ as before and using elementary rules for derivatives as well as the symmetry of the standard normal PDF $\varphi$, one gets
\begin{scriptsize}
\begin{align*}
    &\frac{\partial \ell}{\partial \eta_{\sigma_k}}=\frac{\partial \sigma_k}{\partial \eta_{\sigma_k}}\frac{\partial \ell}{\partial \sigma_k}=\sigma_k\Bigg\{\omega_k\left[-2\left(\frac{y-\mu_k}{\sigma_k}\right)^2\varphi\left(\frac{y-\mu_k}{\sigma_k}\right)+2\varphi\left(\frac{y-\mu_k}{\sigma_k}\right)+2\left(\frac{y-\mu_k}{\sigma_k}\right)^2\varphi\left(\frac{y-\mu_k}{\sigma_k}\right)\right]-\sqrt{2}\varphi(0)\omega_k^2\\
    &-\sum\limits_{l=1}^{k-1}\omega_k\omega_l\Bigg[-2\sigma_k\frac{(\mu_l-\mu_k)^2}{(\sigma_l^2+\sigma_k^2)^{\frac{3}{2}}}\varphi\left(\frac{\mu_l-\mu_k}{\sqrt{\sigma_l^2+\sigma_k^2}}\right)+2\frac{\sigma_k}{\sqrt{\sigma_l^2+\sigma_k^2}}\varphi\left(\frac{\mu_l-\mu_k}{\sqrt{\sigma_l^2+\sigma_k^2}}\right)+2\sigma_k\frac{(\mu_l-\mu_k)^2}{(\sigma_l^2+\sigma_k^2)^{\frac{3}{2}}}\varphi\left(\frac{\mu_l-\mu_k}{\sqrt{\sigma_l^2+\sigma_k^2}}\right)\Bigg]\\
    &-\sum\limits_{l=k+1}^{K}\omega_l\omega_k\Bigg[-2\sigma_k\frac{(\mu_k-\mu_l)^2}{(\sigma_k^2+\sigma_l^2)^{\frac{3}{2}}}\varphi\left(\frac{\mu_k-\mu_l}{\sqrt{\sigma_k^2+\sigma_l^2}}\right)+2\frac{\sigma_k}{\sqrt{\sigma_k^2+\sigma_l^2}}\varphi\left(\frac{\mu_k-\mu_l}{\sqrt{\sigma_k^2+\sigma_l^2}}\right)+2\sigma_k\frac{(\mu_k-\mu_l)^2}{(\sigma_k^2+\sigma_l^2)^{\frac{3}{2}}}\varphi\left(\frac{\mu_k-\mu_l}{\sqrt{\sigma_k^2+\sigma_l^2}}\right)\Bigg]\Bigg\}\\  
    &=2\omega_k\sigma_k\varphi\left(\frac{y-\mu_k}{\sigma_k}\right)
    -\sqrt{2}\varphi(0)\omega_k^2\sigma_k
    -\omega_k\sum\limits_{l=1}^{k-1}\omega_l\frac{2\sigma_k^2}{\sqrt{\sigma_l^2+\sigma_k^2}}\varphi\left(\frac{\mu_l-\mu_k}{\sqrt{\sigma_l^2+
    \sigma_k^2}}\right)
    -\omega_k\sum\limits_{l=k+1}^{K}\omega_l\frac{2\sigma_k^2}{\sqrt{\sigma_k^2+\sigma_l^2}}\varphi\left(\frac{\mu_k-\mu_l}{\sqrt{\sigma_k^2+
    \sigma_l^2}}\right)\\
    &=2\omega_k\sigma_k\left[\varphi\left(\frac{y-\mu_k}{\sigma_k}\right)-\sum\limits_{i=1}^{K}\omega_i\frac{\sigma_k}{\sqrt{\sigma_k^2+\sigma_i^2}}\varphi\left(\frac{\mu_k-\mu_i}{\sqrt{\sigma_k^2+
    \sigma_i^2}}\right)\right].
\end{align*}  
\end{scriptsize}Using the result of Equation \eqref{eq: grad_weight1}, one obtains for the first and second partial sum in Equation \eqref{eq: CRPS3}
\begin{footnotesize}
\begin{align*}
    \frac{\partial}{\partial \eta_{\omega_k}}\sum\limits_{i=1}^{K}\omega_iA(y-\mu_i, \sigma_i^2)&=\sum\limits_{i=1}^{K}(\delta_{ik}\omega_i-\omega_i\omega_k)A(y-\mu_i, \sigma_i^2)=\omega_kA(y-\mu_k, \sigma_k^2)-\omega_k\sum\limits_{i=1}^{K}\omega_iA(y-\mu_i, \sigma_i^2),\\
    \frac{\partial}{\partial \eta_{\omega_k}}\left(-\frac{1}{2}\sum\limits_{i=1}^{K}\omega_i^2A(0, 2\sigma_i^2)\right)&=-\sum\limits_{i=1}^{K}\omega_i(\delta_{ik}\omega_i-\omega_i\omega_k)A(0, 2\sigma_i^2)=-\omega_k^2A(0,2\sigma_k^2)+\omega_k\sum\limits_{i=1}^{K}\omega_i^2A(0, 2\sigma_i^2).
\end{align*} 
\end{footnotesize}For the next step, we split up the third partial sum in Equation \eqref{eq: CRPS3} using property $(\ast)$ to 
\begin{footnotesize}
\begin{align*}
-\sum\limits_{1\leq i< j\leq K}\omega_i\omega_jA(\mu_i-\mu_j, \sigma_i^2+\sigma_j^2)&=-\sum\limits_{i\in \{1,\ldots,K\}\backslash\{k\}}^{}\omega_i\omega_kA(\mu_i-\mu_k, \sigma_i^2+\sigma_k^2)\\
&-\sum\limits_{\substack{i,j\in \{1,\ldots,K\}\backslash\{k\},\\ i<j}}^{}\omega_i\omega_jA(\mu_i-\mu_j, \sigma_i^2+\sigma_j^2).
\end{align*} 
\end{footnotesize}Using again the result of Equation \eqref{eq: grad_weight1} yields for $i\neq k$ into
\begin{footnotesize}
\begin{align}
    \frac{\partial}{\partial \eta_{\omega_k}}\omega_i\omega_k&=\left(\frac{\partial}{\partial \eta_{\omega_k}}\omega_i\right)\omega_k+\omega_i\left(\frac{\partial}{\partial \eta_{\omega_k}}\omega_k\right)=(-\omega_i\omega_k)\omega_k + \omega_i(\omega_k-\omega_k^2)=\omega_i\omega_k(1-2\omega_k), \label{eq: grad_weight2}
\end{align} 
\end{footnotesize}
and for $i<j$ with $i,j\neq k$ to
\begin{footnotesize}
\begin{align}
    \frac{\partial}{\partial \eta_{\omega_k}}\omega_i\omega_j&=\left(\frac{\partial}{\partial \eta_{\omega_k}}\omega_i\right)\omega_j+\omega_i\left(\frac{\partial}{\partial \eta_{\omega_k}}\omega_j\right)=(-\omega_i\omega_k)\omega_j+\omega_j(-\omega_j\omega_k)=-2\omega_i\omega_j\omega_k. \label{eq: grad_weight3}
\end{align} 
\end{footnotesize}
Employing Equations \eqref{eq: grad_weight1}, \eqref{eq: grad_weight2} and \eqref{eq: grad_weight3} allows to calculate 
\begin{footnotesize}
\begin{align*}
    &\frac{\partial}{\partial \eta_{\omega_k}}\sum\limits_{i\in \{1,\ldots,K\}\backslash\{k\}}^{}-\omega_i\omega_kA(\mu_i-\mu_k, \sigma_i^2+\sigma_k^2)=(2\omega_k-1)\sum\limits_{i\in \{1,\ldots,K\}\backslash\{k\}}^{}\omega_i\omega_kA(\mu_i-\mu_k, \sigma_i^2+\sigma_k^2),\\
    &\frac{\partial}{\partial \eta_{\omega_k}}\sum\limits_{\substack{i,j\in \{1,\ldots,K\}\backslash\{k\},\\ i<j}}^{}-\omega_i\omega_jA(\mu_i-\mu_j, \sigma_i^2+\sigma_j^2)=2\omega_k\sum\limits_{\substack{i,j\in \{1,\ldots,K\}\backslash\{k\},\\ i<j}}^{}\omega_i\omega_jA(\mu_i-\mu_j, \sigma_i^2+\sigma_j^2).
\end{align*} 
\end{footnotesize}Putting all results together and reordering the individual terms yields
\begin{footnotesize}
\begin{align*}
    \frac{\partial \ell}{\partial \eta_{\omega_k}}&=\omega_kA(y-\mu_k, \sigma_k^2)-\omega_k\sum\limits_{i=1}^{K}\omega_iA(y-\mu_i, \sigma_i^2)-\omega_k^2A(0,2\sigma_k^2)+\omega_k\sum\limits_{i=1}^{K}\omega_i^2A(0, 2\sigma_i^2)\\
    &+(2\omega_k-1)\sum\limits_{i\in \{1,\ldots,K\}\backslash\{k\}}^{}\omega_i\omega_kA(\mu_i-\mu_k, \sigma_i^2+\sigma_k^2)+2\omega_k\sum\limits_{\substack{i,j\in \{1,\ldots,K\}\backslash\{k\},\\ i<j}}^{}\omega_i\omega_jA(\mu_i-\mu_j, \sigma_i^2+\sigma_j^2)\\
    &=\omega_k\sum\limits_{i=1}^{K}\omega_iA(y-\mu_i, \sigma_i^2)+\omega_kA(y-\mu_k, \sigma_k^2)-\omega_k^2A(0,2\sigma_k^2)-\sum\limits_{i\in \{1,\ldots,K\}\backslash\{k\}}^{}\omega_i\omega_kA(\mu_i-\mu_k, \sigma_i^2+\sigma_k^2)\\
    &-2\omega_k\underbrace{\left[\sum\limits_{i=1}^{K}\omega_iA(y-\mu_i, \sigma_i^2)-\frac{1}{2}\sum\limits_{i=1}^{K}\omega_i^2A(0,2\sigma_i^2)-\sum\limits_{1\leq i< j\leq K}\omega_i\omega_jA(\mu_i-\mu_j, \sigma_i^2+\sigma_j^2)\right]}_{=\mathrm{CRPS}(F,y)}\\
    &=-2\omega_k\mathrm{CRPS}(F,y)+\omega_k\sum\limits_{i=1}^{K}\omega_iA(y-\mu_i, \sigma_i^2)+\omega_kA(y-\mu_k, \sigma_k^2)-\omega_k\sum\limits_{i=1}^{K}\omega_iA(\mu_i-\mu_k, \sigma_i^2+\sigma_k^2)\\
    &=-2\omega_k\mathrm{CRPS}(F,y)+\sum\limits_{i=1}^{K}(\delta_{ik}\omega_i-\omega_i\omega_k)A(y-\mu_i, \sigma_i^2)-\omega_k\sum\limits_{i=1}^{K}\omega_iA(\mu_i-\mu_k, \sigma_i^2+\sigma_k^2).
\end{align*}
\end{footnotesize}

\begin{flushright}
$\Box$
\end{flushright}

\newpage
\addcontentsline{toc}{section}{References}
\thispagestyle{plain}
\clearpage

\printbibliography

\end{document}